\documentclass[12pt]{iopart}
\usepackage{epsfig}
\usepackage{iopams}
\usepackage{cite}

\def\eqspace{\hspace{2.0cm}}
\def\PRE{{\it Phys. Rev. E} }
\def\JSP{{\it J. Stat. Phys.} }
\def\half{\frac{1}{2}}
\def\av#1{\left< #1 \right>}

\def\ie{{\it i.e.} }

\begin{document}

\title[SSB in a non-conserving model]{Spontaneous Symmetry Breaking in a Non-Conserving Two-Species Driven Model}

\author{E. Levine $\dagger$ and R. D. Willmann $\ddagger$}

\address{$\dagger$ Department of Physics of Complex Systems, Weizmann
Institute of Science, Rehovot, Israel 76100.}
\address{$\ddagger$ Institut f\"ur Festk\"orperforschung, Forschungszentrum
J\"ulich, 52425 J\"ulich, Germany}

\begin{abstract}
A two species particle model on an open chain with dynamics which
is non-conserving in the bulk is introduced. The dynamical rules
which define the model obey a symmetry between the two species.
The model exhibits a rich behavior which includes spontaneous
symmetry breaking and localized shocks.
The phase diagram in several regions of parameter space is
calculated within mean-field approximation, and compared with
Monte-Carlo simulations. In the limit where fluctuations in the
number of particles in the system are taken to zero, an exact
solution is obtained. We present and analyze a physical picture
which serves to explain the different phases of the model.
\end{abstract}



\section{Introduction}
\label{sec:intro} One dimensional driven diffusive systems have
been shown in the last decade to demonstrate a variety of
non-trivial types of behavior. In contrast to equilibrium systems
with local interactions, driven systems were shown to exhibit
boundary induced phase transitions, phase separation and
spontaneous symmetry breaking even when the dynamics is local
\cite{Zia,Mukamel00,Evans00,Schutz01,Schutz03}. The latter
phenomenon was first demonstrated for a driven two-species model
with open boundary conditions, which became known as the `Bridge
model'
\cite{Evans95,Mukamel00}. 
This model is defined on an open lattice, where two species of
particles, which can be thought of as carrying opposite `charges',
are moving in opposite directions. Although the dynamical rules
which define the model are symmetric under charge exchange and
left-right reflection, a mean-field approach and Monte Carlo
simulations showed the existence of two phases in which this
symmetry is broken. This result was subsequently supported by an
exact solution for the limit of small extraction rates at the
boundaries \cite{Godreche95}.

Until recently, studies of one dimensional driven systems focused
on conserving bulk dynamics. Non-conserving dynamics was only
considered at certain sites, namely defect sites or boundaries.
Recently, the existence of phase transitions in driven systems
without bulk particle conservation was demonstrated, both for
finite non-conserving rates \cite{Evans02} and for rates which
scale to zero with the system size \cite{Parmeggiani03,Popkov03B,Evans03}.

In this work we study a variant of the Bridge model, in which
non-conserving dynamics  is introduced in the bulk of the system.
All dynamical rates, both conserving and non-conserving, respect
the CP-symmetry between the two species of particles.  By its
nature, this dynamics acts to balance the
densities of the two species. 
If the non-conserving transitions occur at finite rates,
spontaneous symmetry breaking does not occur. However, we find
that when these rates are inversely proportional to the system
size, spontaneous symmetry breaking appears. This model is
related to the single-species non-conserving asymmetric exclusion
process which was introduced and studied recently
\cite{Parmeggiani03,Popkov03B,Evans03}. As in that model, the
two-species model exhibits new phases with localized shocks.
Transitions in the phase diagram can be understood by considering
the position of the localized shocks. Furthermore, the phenomenon
of induced localized shocks is observed, as predicted in
\cite{Popkov03A}.

The paper is organized as follows. The next section contains the
definition of the model. In section \ref{sec:mft} we
calculate the phase diagram within mean-field approximation, for
some cases where the bulk dynamics of the two species are
decoupled. Different limits of the model are discussed, and in
particular the phase diagram of the Bridge model is recovered as
the non-conserving rates are reduced to zero. Next, we present in
section \ref{sec:mc} results of Monte-Carlo simulations, and
compare them with the predictions of mean-field analysis. The
phenomenon of induced shocks is addressed in section
\ref{sec:induced}. In section \ref{sec:toy} we present an exact
solution for the case where fluctuations in the number of
particles in the system are taken to zero. A physical picture is
introduced and analyzed in section \ref{sec:pic}.
 We conclude and summarize in section
\ref{sec:summary}.

\section{Model Definition}
\label{sec:def} The model considered in this paper is defined on a
one-dimensional lattice of size $N$. Each lattice site can either
be occupied by a positive ($+$) particle, occupied by a negative
($-$) particle, or vacant ($0$). The system evolves through three
types of stochastic rates: In the bulk of the system particles
move on the lattice according to the rates
\begin{equation}
\label{eq:rates1}
+0 \mathop{\to}^1 0+ \eqspace +- \mathop{\to}^q -+ \eqspace 0- \mathop{\to}^1 -0\;.
\end{equation}
In addition, each site in the bulk of the system, $1 < i < N$,
can change its state with rates
\begin{equation}
\label{eq:rates2}
+ \mathop{\rightleftharpoons}^{\omega_X}_{\omega_X} - \eqspace +
\mathop{\rightleftharpoons}^{\omega_D}_{\omega_A} 0 \eqspace -
\mathop{\rightleftharpoons}^{\omega_D}_{\omega_A} 0\;,
\end{equation}
corresponding to charge exchange, desorption of a particle from
the lattice, and adsorption of a particle at an empty site. At the
boundaries particles may be introduced and removed. At the left
boundary, site $i=1$, positive particles are introduced and
negative particles are removed with rates
\begin{equation}
\label{eq:rates3}
0 \mathop{\to}^\alpha + \eqspace - \mathop{\to}^\beta 0 \eqspace (i=1)\;,
\end{equation}
while at the right boundary, $i=N$, negative particles are
introduced and positive particles are removed with rates
\begin{equation}
\label{eq:rates4}
0 \mathop{\to}^\alpha - \eqspace + \mathop{\to}^\beta 0\eqspace (i=N)\;.
\end{equation}
Note that all dynamical rules, conserving and non-conserving, are
CP-symmetric, namely symmetric under the exchange of
positive-negative charges and left-right directions.
Generalizations of this model to the case were both types of
particles can move in both directions, and when the dynamical
rules break the CP symmetry, will be considered elsewhere.

Considering the bulk non-conserving rates, one distinguishes three possible
scenarios \cite{Popkov03B}. If the rates are finite, in the thermodynamic limit
the bulk densities are dominated by the non conserving kinetics. Otherwise, if
the rates decay to zero faster than $1/N$, the bulk non-conservation should
have no effect, and the properties of the system in the thermodynamic limit
are identical to those with bulk conservation. The third case, which we
consider here, is the one in which the non-conserving rates scale down
linearly with the system size. It is useful to introduce the
notation  $\omega_A = \Omega/N$, $\omega_X = \Omega u_X/N$, $\omega_D = \Omega u_D/N$.

Without the non-conserving dynamics in the bulk of the system, Eq.
(\ref{eq:rates2}), the model is identical to the Bridge model
introduced in \cite{Evans95} and further studied in
\cite{Godreche95,Arndt98,Clincy01}. Let us summarize first the
known results for this model. At $q=1$ mean-field analysis
predicts the following phase diagram \cite{Evans95}. Two phases
which obey the CP-symmetry of the model are identified: a
maximal-current phase, where the bulk density of each species
equals $\half$, and a low-density phase, where the bulk densities
of both species are identical, both smaller than $\half$. In
addition, the model exhibits two phases in which the symmetries
are broken. In one phase, termed the Low-High phase, one of the
species, spontaneously chosen, sustains a bulk density
 larger than $\half$, while the bulk
density of the other is smaller than $\half$. In the other phase
both species sustain low albeit non-equal densities. This phase
spans a very small area of the mean-field phase diagram, and is
very difficult to recover in Monte-Carlo simulations.
Consequently, the existence of this low-density asymmetric phase
under noisy dynamics has been the subject of some debate
\cite{Arndt98,Clincy01}.

The model considered here can be thought of as two interacting
single-species totally asymmetric exclusion processes with bulk
non-conservation (NC-TASEP). The NC-TASEP is an extension of the
well-known totally asymmetric exclusion process (TASEP)
\cite{TASEP} with bulk absorption and desorption dynamics. Using
mean-field calculations \cite{Evans03}, which were argued to be
exact \cite{Popkov03B}, the phase diagram of this model was
obtained. In addition to the three phases of the TASEP, namely the
maximal-current phase, the high-density phase and the low-density
phase, it was found that the NC-TASEP may also exhibit four
additional phases. The most interesting one is a Shock phase which
consists of a localized shock in the bulk of the system,
separating a low-density region to its left from a high-density
region to its right. 
In the bulk-conserving TASEP shocks appear only on the boundary line between the high-density and low-density phases. On that line a delocalized shock appears in the system. As the
position of the shock is equally probable at any site in the
system, the average profile on the transition line is linear. 
In contrast, the NC-TASEP exhibits a distinct phase in which a localized
shock, whose position is selected by the dynamics, is present. The existence
of this phase plays a main role in our analysis of the two-species
model.

The maximal-current phase of the TASEP appears in the NC-TASEP
only when the two non-conserving rates, namely particle absorption
and desorption, are equal. In this case there exist three more
phases. These include a Low-Max phase, in which the density
increases linearly from a boundary density $<\half$ towards
density $\half$, where it remains constant for the rest of the
system ; a Max-High phase,  in which the density rises linearly
from a constant profile of density $\half$ to a  boundary density
$>\half$ ; and a Low-Max-High phase, in which the density rises
linearly  from a left-boundary density $<\half$ towards $\half$,
where it remains constant up to a point where it climbs linearly
again towards a right-boundary density $>\half$.

\section{Mean Field Theory}
\label{sec:mft} In this section we study the mean-field equations
of our model in the thermodynamic limit $N \to \infty$. We
introduce the occupation variables $\tau_i$ and $\theta_i$, such
that $\tau_i=1$ ($\theta_i=1$) if site $i$ is occupied by a
positive (negative) particle, and $0$ otherwise. The densities of
the positive and negative particles are then defined by
\begin{equation}
p_i = \av{\tau_i}\eqspace m_i = \av{\theta_i}\;,
\end{equation}
where angular brackets denote averaging over realizations.

The time evolution of the particle densities is governed by
\begin{eqnarray}
\label{eq:evolution}
\frac{d p_i}{dt} &=& j^+_{i-1}-j^+_{i}+S^+_i \nonumber\\
\frac{d m_i}{dt} &=& j^-_{i+1}-j^-_{i}+S^-_i \;,
\end{eqnarray}
where the currents are given by
\begin{eqnarray}
\label{eq:bulk}
j^+_i &=& \av{\tau_i(1-\tau_{i+1}-(1-q)\theta_{i+1})} \nonumber\\
j^-_i &=& \av{\theta_i(1-\theta_{i-1}-(1-q)\tau_{i-1})} \;,
\end{eqnarray}
and the source terms are
\begin{eqnarray}
\label{eq:source}
S^+_i &=& \omega_A\left(1-p_i-m_i\right) - \omega_Dp_i + \omega_X\left(m_i-p_i\right)\nonumber \\
S^-_i &=& \omega_A\left(1-m_i-p_i\right) - \omega_Dm_i + \omega_X\left(p_i-m_i\right)\;.
\end{eqnarray}
At the boundaries the source terms vanish, and the currents are
given by
\begin{eqnarray}
\label{eq:mfboundaries}
j^{+}_0 &=& \alpha\left(1-p_1-m_1\right) \nonumber \\
j^{+}_N &=& \beta p_N \\
j^{-}_1 &=& \beta m_1 \nonumber \\
j^{-}_{N+1} &=& \alpha\left(1-p_N-m_N\right) \nonumber \;.
\end{eqnarray}
The mean-field theory for this model is defined by replacing
two-point correlation functions with products of one-point
averages. Within this approximation, the currents become
\begin{eqnarray}
\label{eq:mfbulk}
j^+_i &=& {p_i\left(1-p_{i+1}-(1-q)m_{i+1}\right)} \nonumber\\
j^-_i &=& {m_i\left(1-m_{i-1}-(1-q)p_{i-1}\right)} \;.
\end{eqnarray}
In the steady-state all time derivatives vanish, and one has
\begin{equation}
j^+_{i} = j^{+}_{i-1}+S_i^{+} \eqspace
j^-_{i} = j^{+}_{i+1}+S_i^{+} \;.
\end{equation}
Defining ${\cal J}^{+}_i = j^+_{i} - \sum_{k=0}^{i}S^{+}_k , {\cal
J}^{-}_i = j^-_{i} - \sum_{k=i}^{N}S^{-}_k$, one notices that in
fact ${\cal J}^{+}_i \equiv {\cal J}^{+}$ and ${\cal J}^{-}_i
\equiv {\cal J}^{-} $ are conserved throughout the lattice, and
${\cal J}^{+} = j^{+}_0 , {\cal J}^{-} = j^{-}_{N+1}$.

\subsection{Solution of the mean-field equations in the bulk-decoupled case}
\label{sec:wd0}
The case $q=1, u_X=1$ is special, as in this case the
bulk equations (\ref{eq:evolution}) with the mean-field currents (\ref{eq:mfbulk}) are
decoupled. The currents and source terms in this case are just
\begin{equation}\begin{array}{rclrcl}
j^{+}_{i} &=& p_i\left(1-p_{i+1}\right) & \eqspace S^{+}_i &=& \frac{\Omega}{N}\left[1-(2+u_D)p_i\right] \nonumber\\
j^{-}_{i} &=& m_i\left(1-m_{i-1}\right) & \eqspace S^{-}_i &=&
\frac{\Omega}{N}\left[1-(2+u_D)m_i\right]\;.
\end{array}\end{equation}
In the bulk of the system, the hopping rates for say a positive
particle do not distinguish between a negative particle and a
vacancy. Also, the fact that attachment of a positive particle is
limited by the presence of negative ones, is exactly compensated
by the charge exchange rate. The coupling between the two species
is limited in this case only to the boundaries. Following
\cite{Evans95}, one readily notices that upon the definition
\begin{eqnarray}
\label{eq:effective}
\alpha^+ &=& \frac{\alpha(1-p_1-m_1)}{1-p_1} = \frac{\cal{J}^+}{\frac{{\cal J}^+}{\alpha}+\frac{{\cal J}^-+{\cal S}^-}{\beta}} = \frac{j^{+}_0}{\frac{j_0^+}{\alpha}+\frac{j^-_0}{\beta}}\nonumber \\
\alpha^- &=& \frac{\alpha(1-p_N-m_N)}{1-m_N} = \frac{\cal{J}^-}{\frac{{\cal J}^-}{\alpha}+\frac{{\cal J}^++{\cal S}^+}{\beta}}= \frac{j^{-}_0}{\frac{j_N^-}{\alpha}+\frac{j^+_N}{\beta}} \;,
\end{eqnarray}
with  $S^{\pm} = \sum_{i=1}^{N}S^{\pm}_i$, the problem is reduced
to two single-species NC-TASEP. One process corresponds to the
positive particles with injection rate $\alpha^+$ at the left
boundary and ejection rate $\beta$ at the right, and the other to
the negative particles with injection rate $\alpha^-$ at the right
boundary and ejection rate $\beta$ at the left. The two processes
may or may not share the same phase. The latter case corresponds
to a phase of the two-species system, where the symmetry between
the two species is broken. In the other case, it may be that the
average densities of the two species are not equal, although the
two lie in the same phase of the corresponding NC-TASEP. A trivial
restriction on the possible phases in the model is $p_i+m_i \leq
1$ at all sites. This immediately excludes several possibilities,
such as ones which mix the high-density phase of one species with
anything but the low-density of the other.

In this section we explore the possible phases in the bulk
decoupled case. Symmetric phases are presented first, followed by
asymmetric phases.
In the symmetric phases, $\alpha^+=\alpha^-$, so
for these phases only $\alpha^+$ is quoted in the following. For
the asymmetric phases it is always assumed, with no loss of
generality, that the positive particles are in the majority. When
describing density profiles we always take a language in which the
lattice is rescaled to the segment $[0,1]$. The emerging phase
diagram is discussed in the following subsection.

\begin{figure}[t]
\begin{flushleft}
\includegraphics*[angle=270,width=5cm]{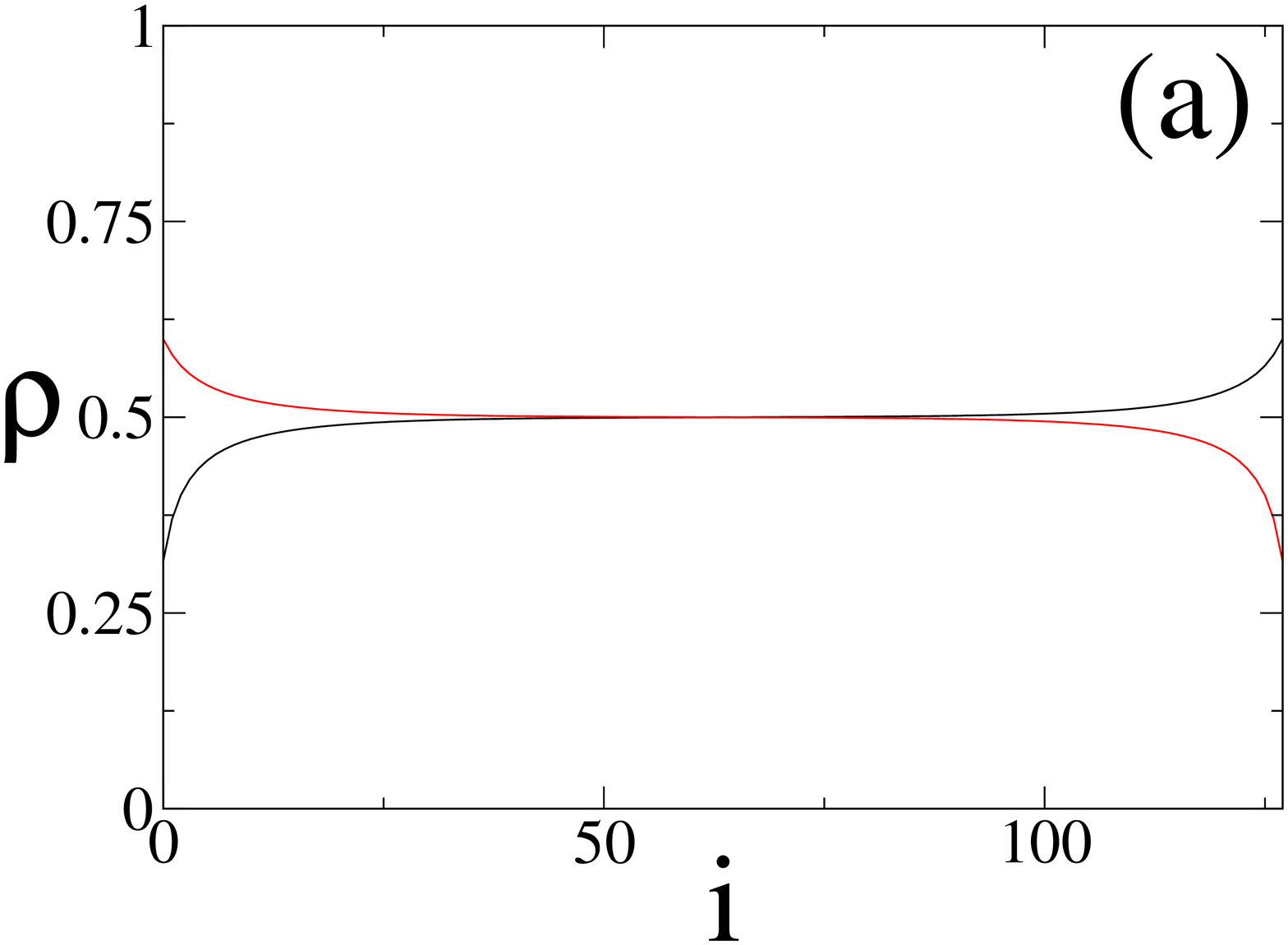}
\includegraphics*[angle=270,width=5cm]{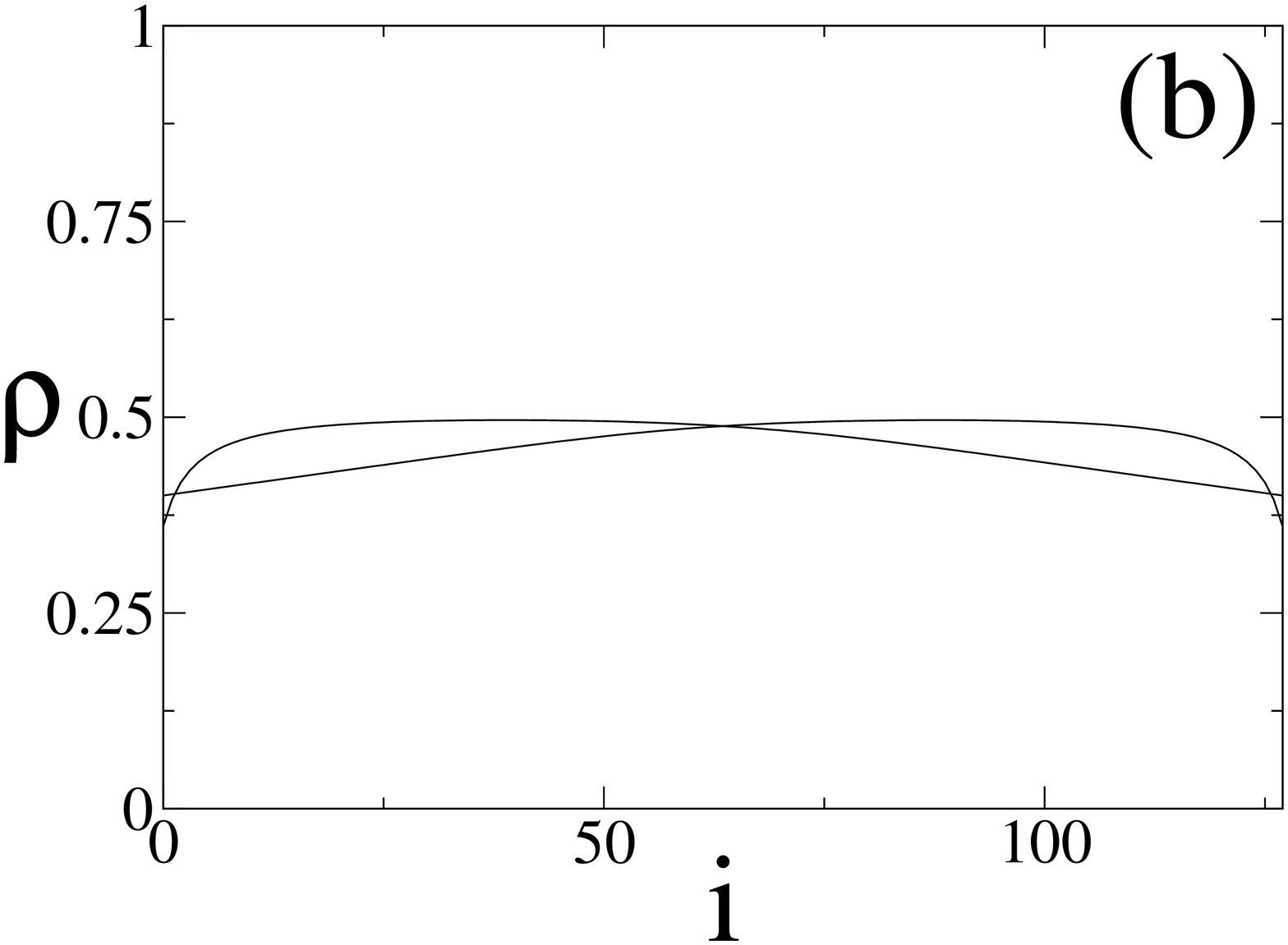}
\includegraphics*[angle=270,width=5cm]{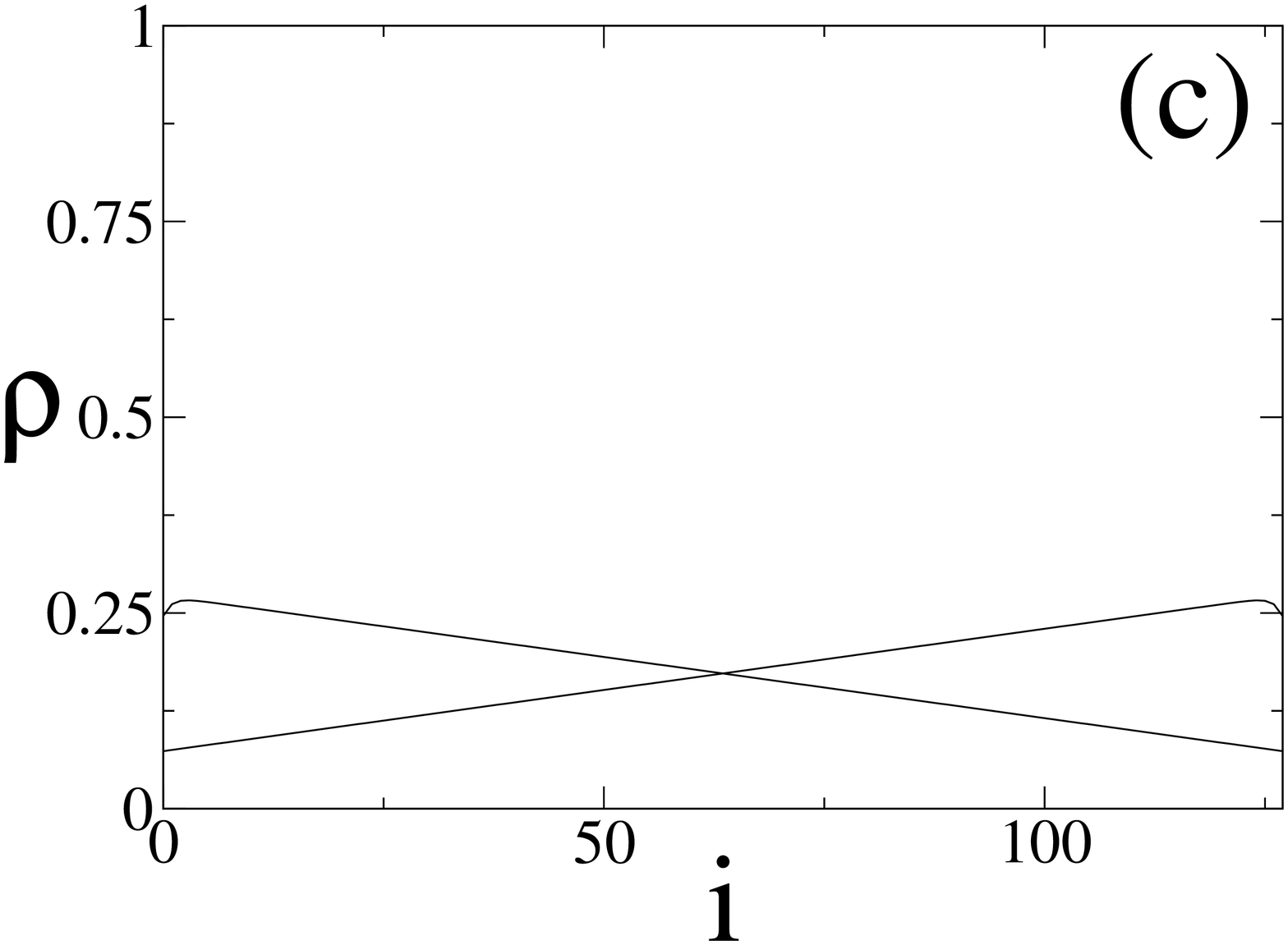}
\\
\includegraphics*[angle=270,width=5cm]{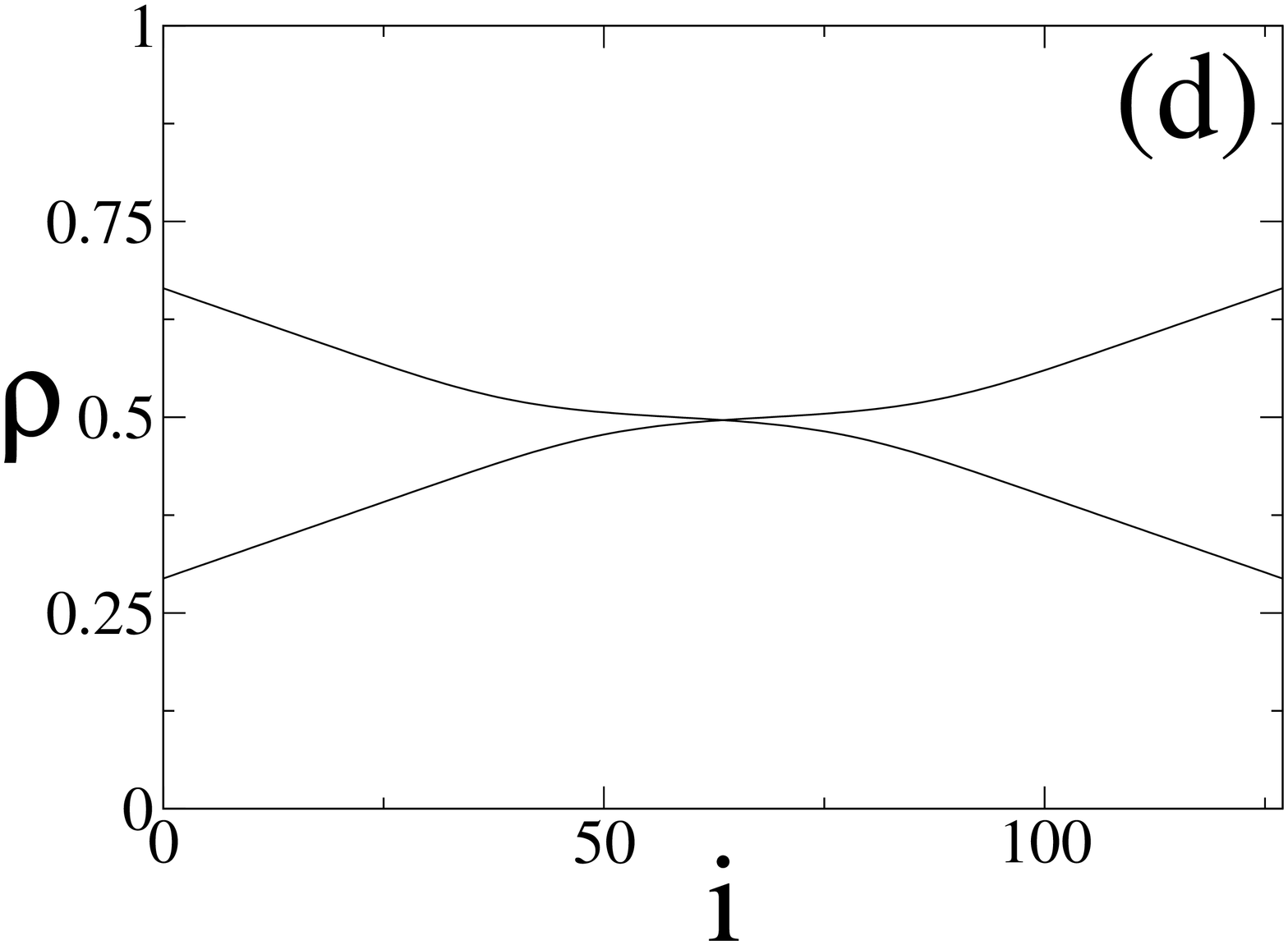}
\includegraphics*[angle=270,width=5cm]{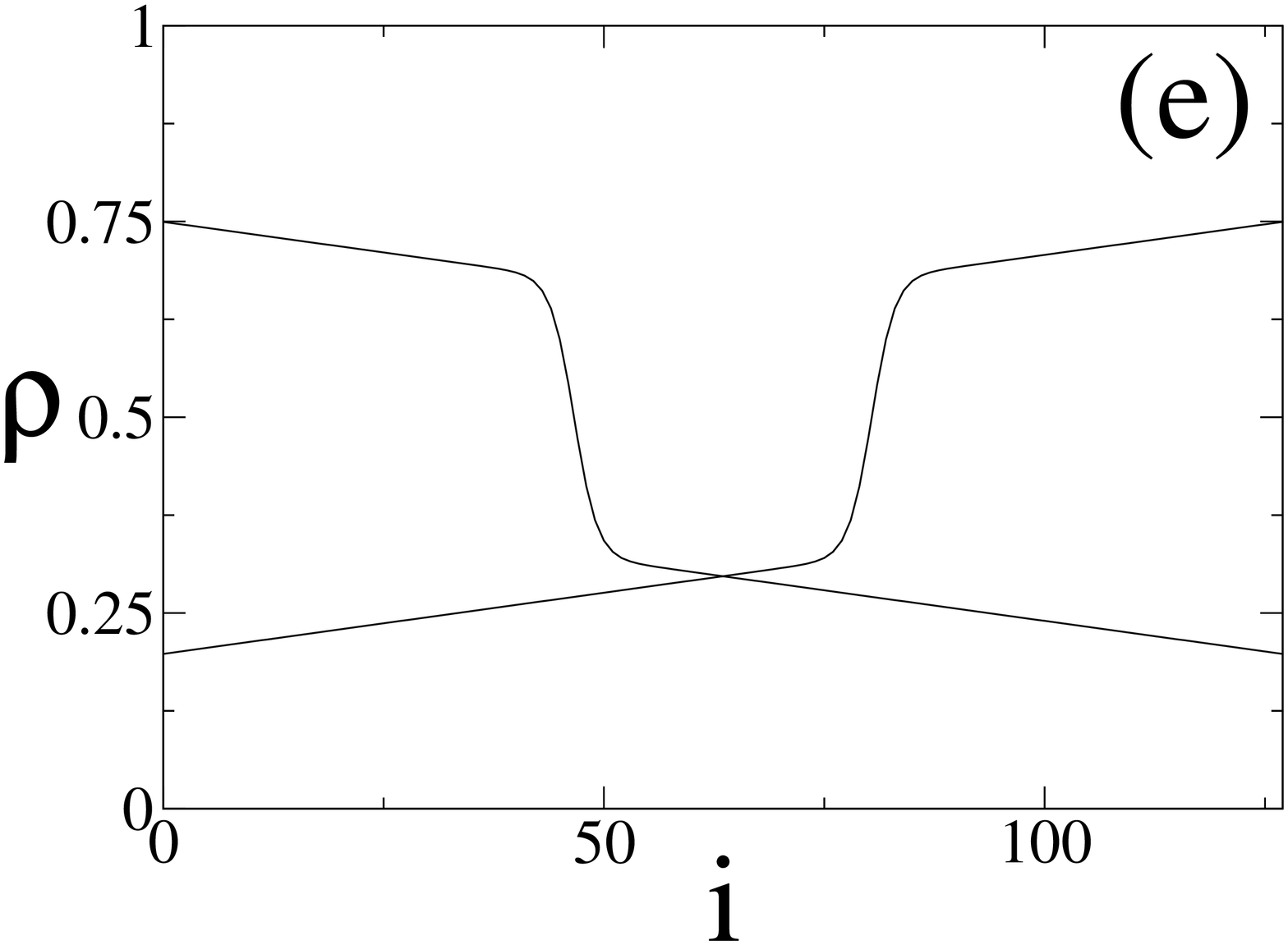}
\caption{Density profiles of the symmetric phases, as obtained by
integrating the mean-field equations for a system of size
$N=128$. (a) Max Phase, $\alpha=3.0$, $\beta=0.8$, $\Omega=0.2$,
$q=1$. (b) Low-Max Phase, $\alpha=1.0$, $\beta=0.7$, $\Omega=0.2$,
$q=1$. (c) Low Phase, $\alpha=0.1$, $\beta=0.8$, $\Omega=0.2$,
$q=1$. (d) Low-Max-High Phase, $\alpha=5.0$, $\beta=1/3$,
$\Omega=0.5$, $q=1$. (e) Shock Phase, $\alpha=3.0$, $\beta=0.25$,
$\Omega=0.2$, $q=1$. \label{fig:MFsymm} }
\end{flushleft}
\end{figure}

\subsubsection{Maximal-current symmetric phase.}
In this phase the bulk density of both species is $\half$, and the boundary currents
are given by
\begin{equation}
j_0^{+}=j_0^{-}=\frac{1}{4}\;.
\end{equation}
The conditions for this phase to exist are

\begin{equation}
\alpha^+>\half \eqspace \beta>\half\;.
\end{equation}
Typical density profiles of the two species in this phase, as
obtained from integrating the mean-field equations, are shown in
figure \ref{fig:MFsymm} (a).
\subsubsection{Low - Max symmetric phase.}
The density profile in this phase is composed of a low density
part where the density increases linearly with slope $\Omega$ 
on the rescaled lattice, as
well as a part with constant density $\half$. The boundary
currents are
\begin{equation}
j_0^{+}=\alpha^+(1-\alpha^+) \eqspace j_0^-=\frac{1}{4}\;.
\end{equation}
The conditions for the existence of this phase are
\begin{equation}
\alpha^+<\half \eqspace \alpha^+>\half-\Omega \eqspace \beta<\half\;.
\end{equation}
The density profiles shown in figure \ref{fig:MFsymm} (b) for a
finite system furthermore exhibits a boundary layer, which does
not scale with the system size.
\subsubsection{Low density symmetric phase.}
In this phase both densities remain below $\half$, increasing
throughout the system with constant slope $\Omega$ (figure
\ref{fig:MFsymm}(c)). The boundary currents are given by
\begin{equation}
\label{eq:bcl}
j_0^{+}=\alpha^+(1-\alpha^+) \eqspace j_0^-=(\alpha^++\Omega)(1-\alpha^+-\Omega)\;.
\end{equation}
Necessary conditions for the existence of this phase are
\begin{eqnarray}
\label{eq:condl}
\alpha^+<\beta-\Omega &\mbox{ for }& \beta<\half \nonumber\\
\alpha^+<\half-\Omega &\mbox{ for }& \beta\geq\half\;.
\end{eqnarray}
Inserting the boundary currents (\ref{eq:bcl}) into
(\ref{eq:effective}) yields a quadratic equation for $\alpha^+$.
Using (\ref{eq:condl}) one readily identifies the relevant
solution.
\subsubsection{Low - Max - High symmetric phase.}
The density profiles in this phase are a mixture of three
different pieces - a linear profile of low densities and constant
slope $\Omega$, followed by a flat density profile at density
$\half$, and a linear profile of high-densities of the same slope
(figure \ref{fig:MFsymm}(d)). The boundary currents are now
\begin{equation}
\label{eq:bclmh}
j_0^{+}=\alpha^+(1-\alpha^+) \eqspace j_0^-=\beta(1-\beta)\;.
\end{equation}
This phase region is defined by the conditions
\begin{equation}
\label{eq:condlmh}
\alpha^+>\half-\Omega \eqspace \beta<\half\;.
\end{equation}
Again, one solves the equation for $\alpha^+$ given by
(\ref{eq:effective}) and (\ref{eq:bclmh}), and uses
(\ref{eq:condlmh}) to identify the relevant solution. For a finite
system (as seen in figure \ref{fig:MFsymm}(d)) the transition
between the three parts is not sharp. It only becomes so on the
rescaled lattice as $N \to \infty$.
\subsubsection{Shock symmetric phase.}
This phase is characterized by a localized shock, separating a low-density region from a
high-density one, both of linear profile with slope $\Omega$.
The boundary currents are given by
\begin{equation}
\label{eq:bcs}
j_0^{+}=\alpha^+(1-\alpha^+) \eqspace j_0^-=\beta(1-\beta)\;.
\end{equation}
Notice that $\alpha^+$ in this phase is identical to the one of
the Low-Max-High phase. The Shock symmetric phase is defined by
the conditions
\begin{equation}
\label{eq:conds}
\beta-\Omega<\alpha^+<\half-\Omega\;.
\end{equation}
The position of the shock
$x_s$ is given by
\begin{equation}
\label{eq:shockposition}
x_s = \frac{\beta-\alpha^+}{2\Omega}+\half = \frac{2\beta-(1+\alpha)+\sqrt{(1+\alpha)^2-4\alpha\beta}}{4\Omega}+\half\;.
\end{equation}
In contrast to the thermodynamical limit, the shocks in a finite
system such as in figure \ref{fig:MFsymm} (e) are not sharp.
\begin{figure}[t]
\begin{flushleft}
\includegraphics*[angle=270,width=5cm]{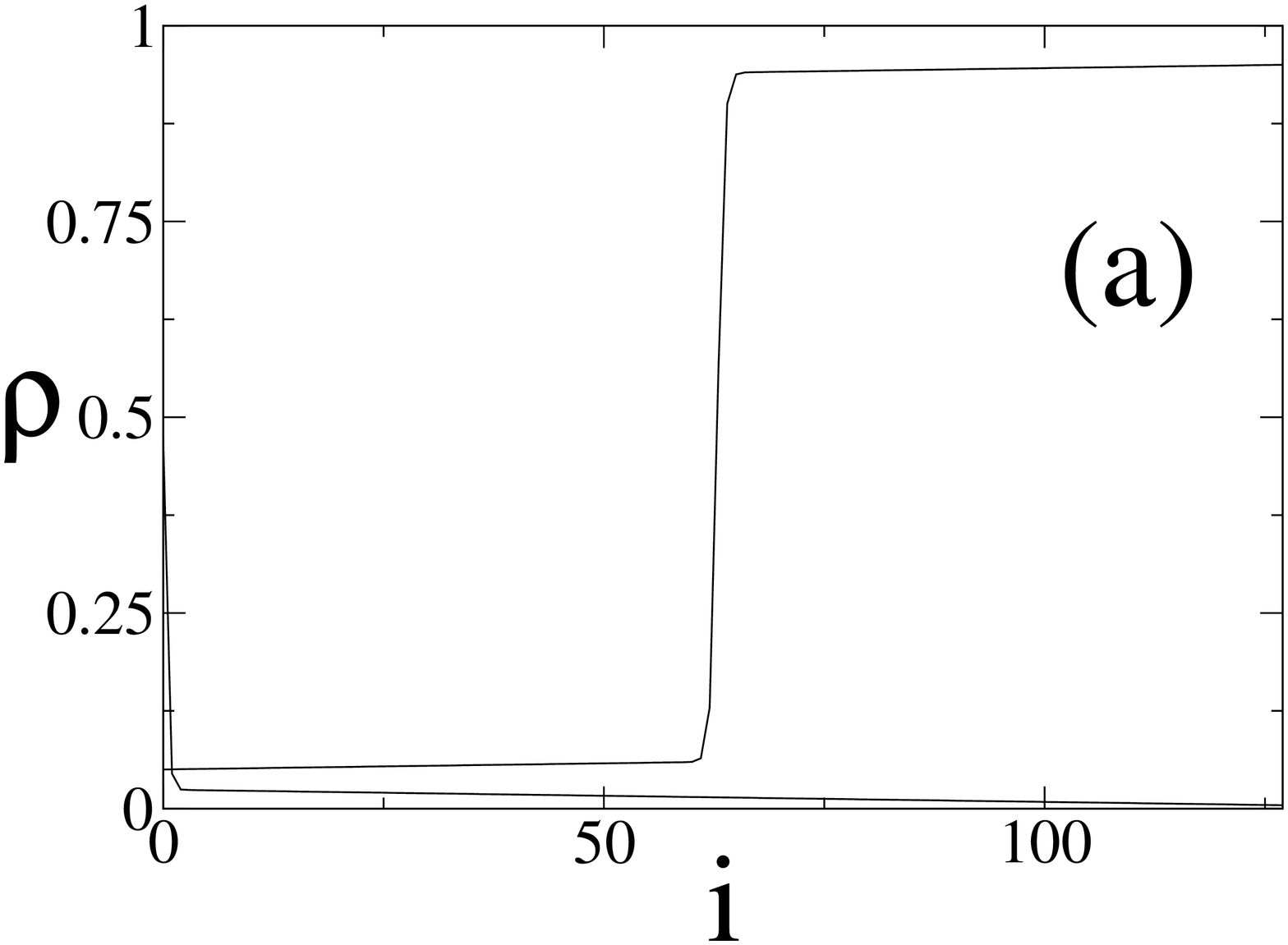}
\includegraphics*[angle=270,width=5cm]{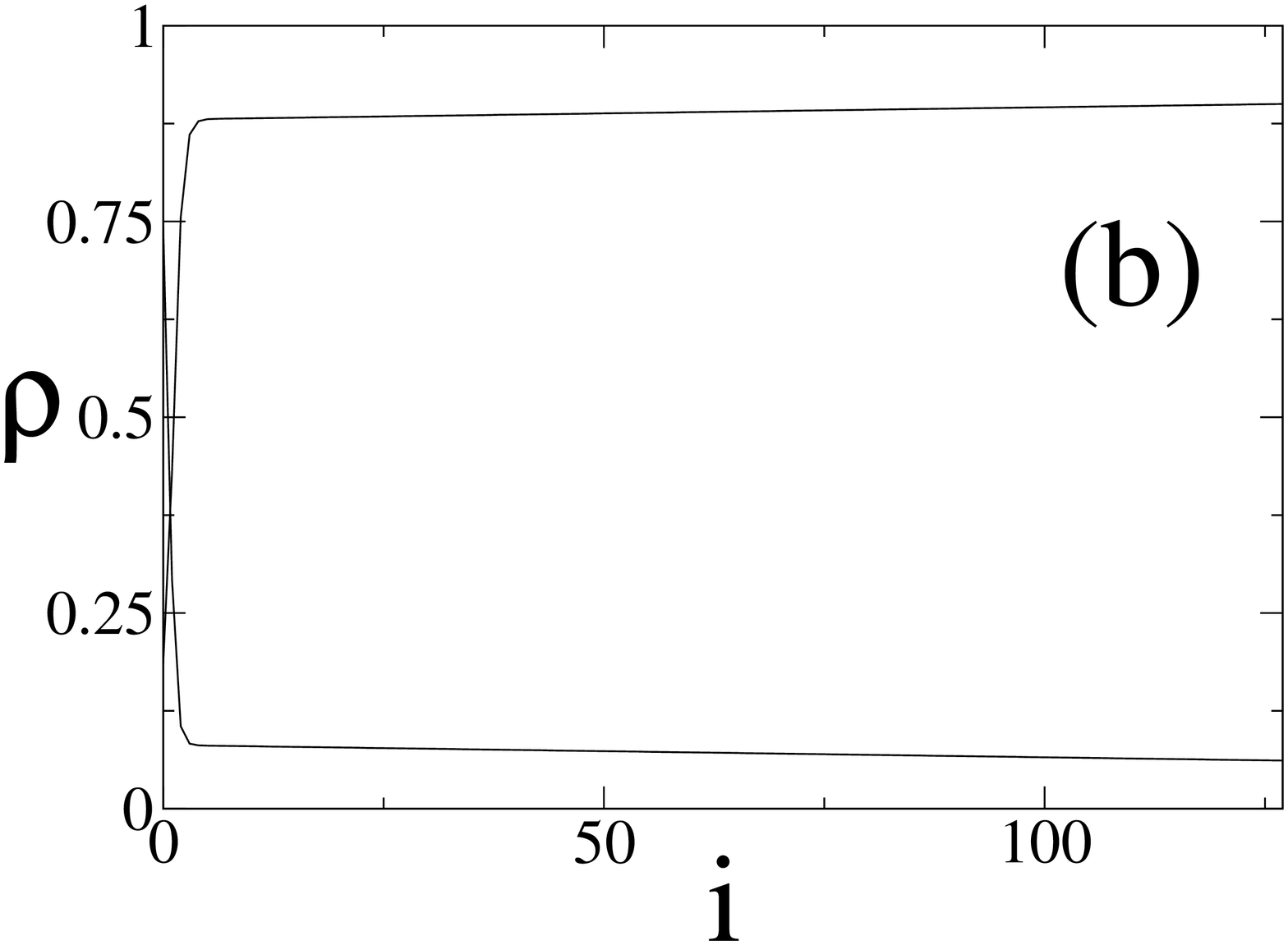}
\includegraphics*[angle=270,width=5cm]{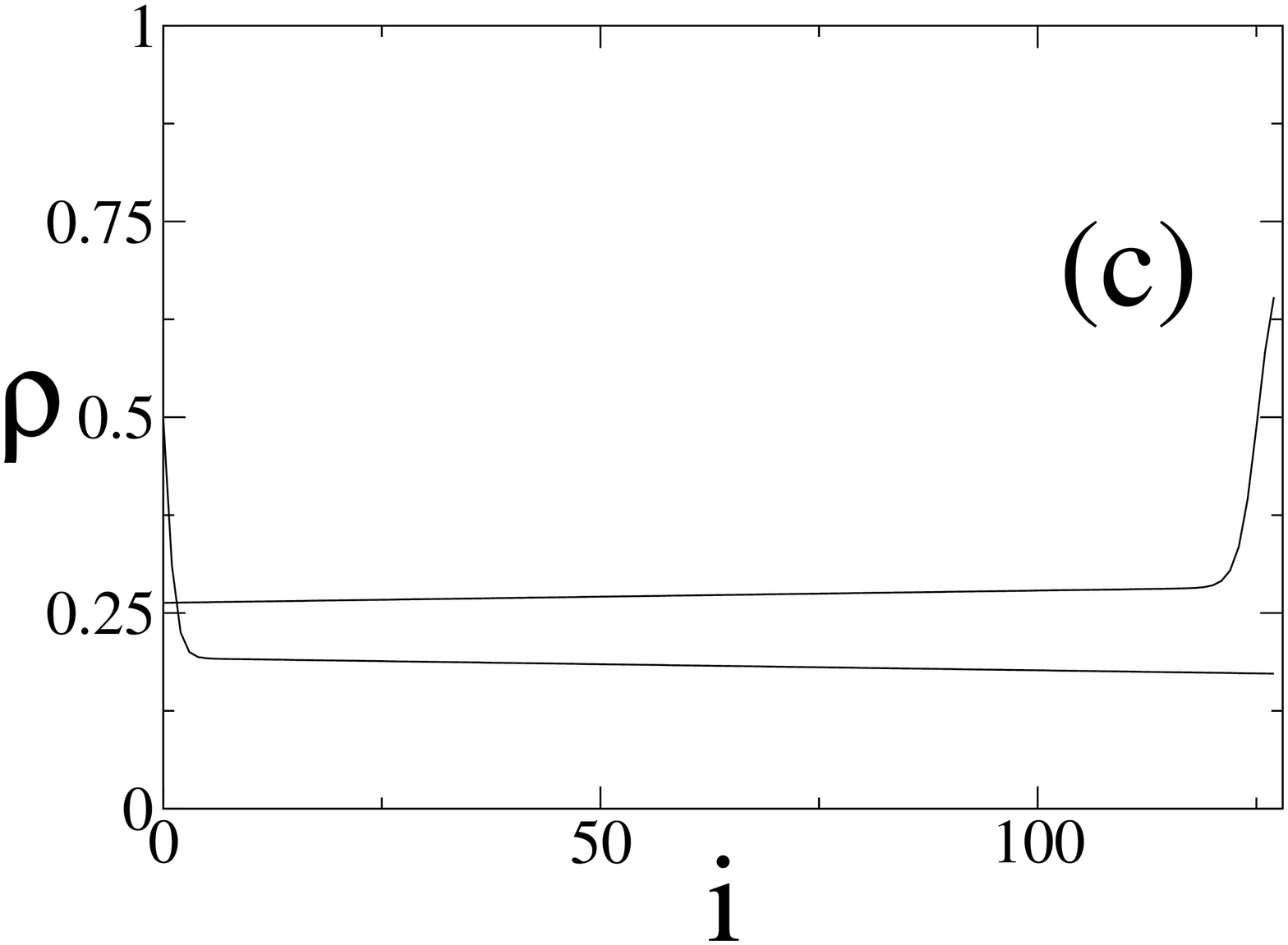}
\caption{Density profiles of the asymmetric phases, as obtained by
integrating the mean-field equations for a system of size $N=128$.
(a) Shock-Low asymmetric Phase, $\alpha=0.1$, $\beta=0.05$,
$\Omega=0.02$, $q=1$. (b) High-Low asymmetric Phase, $\alpha=1.5$,
$\beta=0.1$, $\Omega=0.02$, $q=1$. (c) Low asymmetric Phase,
$\alpha=0.82$, $\beta=0.31$, $\Omega=0.02$, $q=1$.
\label{fig:MFasymm}}
\end{flushleft}
\end{figure}
\subsubsection{Shock - Low asymmetric phase}
In this phase the majority species exhibits a localized shock,
while the minority species is in the Low phase throughout the
system (figure \ref{fig:MFasymm} (a)). The boundary currents for
the two species are
\begin{eqnarray}
\label{eq:bcsl}
j_0^{+}=\alpha^+(1-\alpha^+) &\eqspace&
j_0^-=(\alpha^-+\Omega)(1-\alpha^--\Omega) \nonumber \\
j_N^+=\beta(1-\beta)&\eqspace&
j_N^{-}=\alpha^-(1-\alpha^-) \;.
\end{eqnarray}
The conditions for the existence of this phase are given by
\begin{equation}
\label{eq:condsl}
\alpha^+>\beta-\Omega \eqspace \alpha^+<\beta+\Omega \eqspace \beta<\half\;.
\end{equation}
The equation for $\alpha^-$ does not involve $\alpha^+$, and can
be solved as in previous phases. Plugging this solution into the
equation for $\alpha^+$ one can solve the equation, and identify
the only solution which obeys (\ref{eq:condsl}).
\subsubsection{High - Low asymmetric phase}
This phase is analogous to the strong asymmetric phase of the
Bridge model. In this phase the majority sustains a high density
in the bulk of the system, while the minority density is low.
Here, however, the density profiles are not constant, but rather
of opposite slopes $\pm \Omega$ (figure \ref{fig:MFasymm} (b)).
The boundary currents for the two species are
\begin{eqnarray}
\label{eq:bchl}
j_0^{+}=(\beta+\Omega)(1-\beta-\Omega)&\eqspace&
j_0^-=(\alpha^-+\Omega)(1-\alpha^--\Omega) \nonumber \\
j_N^+=\beta(1-\beta)&\eqspace&
j_N^{-}=\alpha^-(1-\alpha^-) \;.
\end{eqnarray}
The conditions for the existence of this phase are
\begin{equation}
\label{eq:condhl}
\alpha^+>\beta+\Omega \eqspace \beta+\Omega<\half\;.
\end{equation}
Expressions for $\alpha^\pm$ are obtained from (\ref{eq:bchl}),
(\ref{eq:condhl}) just as in the Shock-Low phase.
\subsubsection{Low asymmetric phase}
In this phase both particle species maintain a low density profile
with constant slope $\Omega$. Still, the phase is asymmetric as
the boundary densities of the two phases are different. An
analogous phase is also observed on the mean field level in the
Bridge model. The boundary currents are given by
\begin{eqnarray}
\label{eq:bcll}
j_0^{+}=\alpha^+(1-\alpha^+) &\eqspace& j_0^-=(\alpha^-+\Omega)(1-\alpha^--\Omega)\;, \nonumber \\
j_0^{+}=\alpha^-(1-\alpha^-) &\eqspace& j_0^+=(\alpha^++\Omega)(1-\alpha^+-\Omega)\;.
\end{eqnarray}
Plugging the currents (\ref{eq:bcll}) into (\ref{eq:effective}) gives
\begin{eqnarray}
\label{eq:lleffective}
\alpha^{+} & = &1-\frac{\alpha^{+}(1-\alpha^{+})}{\alpha}-\frac{(\alpha^{-}+\Omega)(1-\alpha^{-}-\Omega)}{\beta} \nonumber \\
\alpha^{-} & =
&1-\frac{\alpha^{-}(1-\alpha^{-})}{\alpha}-\frac{(\alpha^{+}+\Omega)(1-\alpha^{+}-\Omega)}{\beta}\;
\end{eqnarray}
Let $D=\alpha^{+}-\alpha^{-}$ and $S=\alpha^{+}+\alpha^{-}$.
Subtracting the equations in (\ref{eq:lleffective})
yields
\begin{equation}
D=D\left((1-S)\frac{\alpha-\beta}{\alpha \beta}-\frac{2\Omega}{\beta}\right).
\end{equation}
In the asymmetric phase, $D\neq0$, thus an expression for $S$ is
obtained. Summing the equations in (\ref{eq:lleffective}) and
using this result yields $D$ as a function of $\alpha$ and
$\beta$. The effective boundary rates are obtained as
$\alpha^+=\half(S+D)$ and $\alpha^-=\half(S-D)$. Necessary
conditions for the existence of this phase are
\begin{eqnarray}
\label{eq:condll}
\alpha^+<\beta-\Omega &\mbox{ and }& \beta<\half \nonumber\\
\alpha^+<\half-\Omega &\mbox{ and }& \beta\geq\half \nonumber\\
D > 0\;.
\end{eqnarray}
A typical profile for this phase is shown in figure
\ref{fig:MFasymm} (c).
\begin{figure}
\begin{center}\epsfxsize 16 cm \epsfbox{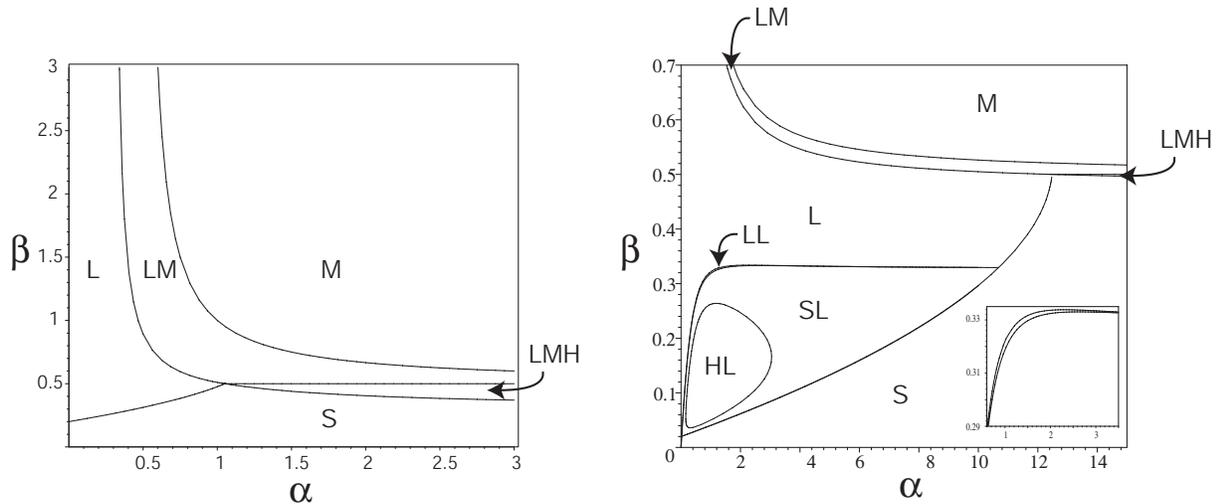}\end{center}
\caption{ Mean-field phase diagram for the
bulk-decoupled case with $\omega_D=0$ and $\Omega = 0.2$ (left),
$\Omega=0.02$ (right). The inset focuses on the regime where the
low-asymmetric phase is most pronounced. Abbreviations used for
the phases: M - maximal-current symmetric phase; LM - Low-Max
symmetric phase; LMH - Low-Max-High symmetric phase; S - Shock
symmetric phase; L - Low symmetric phase; SL - Shock-Low
asymmetric phase; HL - High-Low asymmetric phase; LL - Low
asymmetric phase. }
\label{fig:pd}
\end{figure}

\subsection{Phase Diagram}
In the previous section we have listed all phases which are found to exist in this model, and derived the phase boundaries in which  they reside in parameter space.  We now turn to describe the emerging phase diagram.
First note that the full parameter space is covered by the four
symmetric phases. In fact, all asymmetric phases reside in regions
of phase space where the low-density symmetric state is also
stationary.
Which of the solutions is realized is a matter of stability, as will be discussed shortly. Except for the boundary of the Maximal-current phase, the phase boundaries of the
symmetric phases share a common point $Q$ in the $(\alpha,\beta)$
plane, given by
\begin{equation}
Q=\left( \frac{1-4\Omega^2}{4\Omega}\;,\;\half \right).
\end{equation}
 The intersection point of the phase boundary between the Low and the Shock symmetric
phases meets the $\beta$ axis at the point
\begin{equation}
R=\left(0\;,\;\Omega\right)\;.
\end{equation}
In figure \ref{fig:pd} we plot the phase diagram for the cases
$\Omega=0.2, 0.02$. Note that at the higher value of $\Omega$ no
asymmetric phases exist, while at the smaller value all phases
described in the previous section exist. This fact will be
addressed below. Taking $\Omega$ to zero the original phase
diagram of the Bridge model is retrieved, as discussed in
subsection \ref{sec:omega0}.

\subsubsection{Stability.}
\label{sec:stability}

As mentioned above, all asymmetric phases reside in regions of 
phase space in which the symmetric low density is also a stationary solution of the mean-field equations. On the mean field level, the realization of one stationary solution rather than the other is a matter of initial conditions. In all cases both the symmetric and
asymmetric solutions are linearly stable. However, any initial
condition for which the density of at least one of the two species
is higher than $\half$ in some region evolves into the state of
broken symmetry. Thus a disordered initial condition, in which the
density of particles at any site is an independent uniformly
distributed random variable, resides in the basin of attraction of
the asymmetric solution.

Considering the model 
beyond mean-field approximation, where the dynamics is noisy, one
expects a random perturbation to take the system away from the
symmetric solution. In  physical terms this can be understood by
the fact that nucleation of a high density domain leads to its
flow towards the boundary, where it reduces the inflow of
particles of the other species due to the exclusion interaction.
Once the symmetry is broken, the high density phase expands and
takes its steady state position. This picture is substantiated in
a quantitative manner for the limit $\beta \to 0$ in section
\ref{sec:toy}. Note that this line of argument cannot be followed
for the asymmetric Low phase.

\subsubsection{Extinction of the asymmetric phases at high $\Omega$.}
\label{sec:highomega}
So far we have considered the case $\omega_D=0$, where detachment
of particles from the bulk is suppressed. In this case the
non-conserving rates allow for attachment of particles of either
species with rate $\omega_A$ and for charge exchange with rate
$\omega_X$. While the former process affects the densities of both
species in the same way, the charge exchange process tends to
diminish the density difference between species. Thus it is clear
that for large $\omega_X\sim\Omega$, when this process becomes
dominant, the asymmetric phases will vanish.
This effect can only increase in the presence of detachment, which acts stronger on the majority species.
In the case $\omega_D=0$ we find that the High-Low phase ceases to
exist beyond $\Omega \simeq 0.035$. The Shock-Low phase vanishes
at $\Omega \simeq 0.138$, and with it the Low asymmetric phase.
The vanishing of the asymmetric phases can be understood in a more
quantitative manner from the blockage picture described in section
\ref{sec:pic}.

\subsubsection{The limit $\Omega \to 0$.}
\label{sec:omega0}

In the limit of $\Omega \to 0$ the non-conserving model considered
here must coincide with the Bridge model. In this limit, the point
$R$ defined above is shifted towards the origin. The point $Q$ is
pushed to infinity, which means that the Low-Max-High symmetric
phase cannot exist. The boundary line between the Low symmetric
and the shock symmetric phase is given generally by
\begin{equation}
\beta_{\rm LS}=\Omega+\half-\half\sqrt{1-4\alpha\Omega} \;\mbox{
for }\; \alpha\leq\frac{1-4\Omega^2}{4\Omega}.
\end{equation}
Thus $\beta_{\rm LS} \to 0$ for all $\alpha$ as $\Omega \to 0$,
and the shock symmetric phase cannot exist. Furthermore the
Low-Max symmetric phase vanishes in this limit. The boundary
between this phase and the Maximal-current phase,
\begin{equation}
\beta=\frac{\alpha}{2\alpha-1} \;\mbox{ for }\; \alpha>\half\;,
\end{equation}
coincides in the limit $\Omega \to 0$ with its boundary with the
Low phase,
\begin{equation}
\beta=\frac{\alpha}{2\alpha-1+4\Omega(\alpha+\Omega)} \;\mbox{ for
}\; \half < \alpha<\frac{1-4\Omega^2}{4\Omega}\;.
\end{equation}
Thus, of the symmetric phases only the Maximal-current phase and
the Low phase remain in the $\Omega \to 0$ limit, as expected from
the Bridge model.

As for the asymmetric phases, one notices in the same way that as
$\Omega$ decreases, the High-Low phase region grows on the expense
of the Shock-Low phase. As $\Omega \to 0$, the boundary lines of
both phases coincide, and the Shock-Low phase ceases to exist.
Exact expressions for these phase boundaries are rather lengthy,
and are omitted here. Finally, the low density asymmetric phase
takes in the $\Omega \to 0$ limit its form as in the Bridge model.

\subsection{Detachment from the bulk: the case $\omega_D \neq 0$}
\label{sec:omegad}

\begin{figure}
\begin{center}\hspace{0.05cm}\epsfysize 7 cm \epsfbox{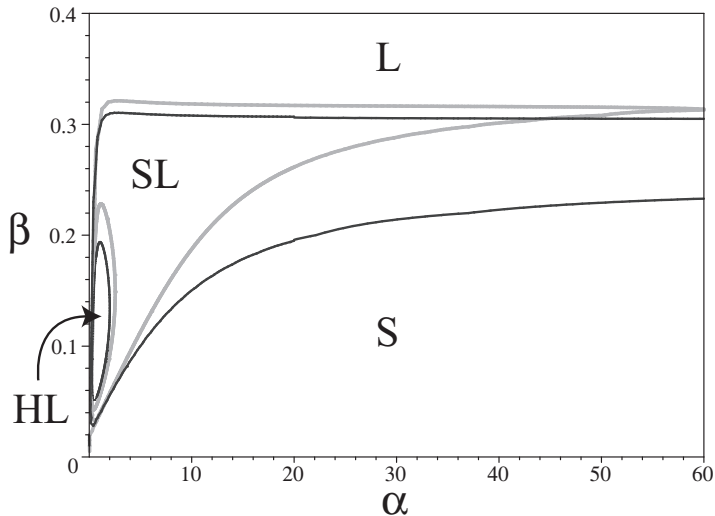} \end{center}
\caption{\label{fig:pdwd} Mean-field phase diagram for the
bulk-decoupled case, with $\Omega = 0.02$ and $u_D=1$ (gray
lines), $u_D = 2$ (dark lines) . The boundary line between the
asymmetric low-density phase and the symmetric Low phase is
omitted. Abbreviations for the different phases are the same as in
figure \ref{fig:pd}. }
\end{figure}

We now turn to consider a more general case, still within the
bulk-decoupled regime ($q=1,\omega_D=\omega_X$), where the
non-conserving dynamics includes detachment of particles. This
case corresponds to the NC-TASEP with non-equal attachment and
detachment rates. The phase diagram of that model includes only
three phases: high density, low density and localized shock. All
phases where a part of the profile is constant at $\half$ do not
exist. This comes from the fact that the equilibrium density of
the kinetics is given by the Langmuir density rather than $\half$.

The phase diagram of our model in the case $\omega_D \neq 0$
exhibits only two symmetric phases - the symmetric low-density
phase and the shock symmetric one. All asymmetric phases which
were obtained for the case $\omega_D = 0$ are also present here.
As in the NC-TASEP, the density profiles in this case are not
linear, but rather curved. This makes the analysis of the phase
diagram somewhat more cumbersome, although not different in
principle from the one presented in section \ref{sec:wd0}. Details
of this calculation are given in the appendix.

In figure \ref{fig:pdwd} we plot the mean-field phase diagram for
the cases $u_D = 1$ and $u_D = 2$,
where $u_D$ is defined as before by $u_D = \omega_D/\omega_A$. The
boundary line between the asymmetric low-density phase and the
symmetric Low phase is not presented. We could not obtain the
boundary densities in this phase in a closed form. For several
values of $u_D$ we have found numerically that this line lies just
above the transition line between the Shock-Low phase and the
asymmetric Low phase. Certainly, the region of phase space covered
by this phase does not increase compared with the case $\omega_D =
0$. It is readily noticed that the part of phase space spanned by
the High-Low asymmetric phase decreases as $\omega_D$ is
increased. This is expected from the fact that the detachment
process acts stronger on the majority phase, thus reducing its
density. For any given $\Omega$ the detachment process can be
increased to a level in which the High-Low asymmetric phase does
not occur.

The detachment process can be considered as cooperating with the
boundary ejection rate $\beta$, and competing with the boundary
injection rate $\alpha$. It is no surprise then that the
asymmetric Shock-Low phase grows in the $\alpha$ direction of
phase space and shrinks in the $\beta$ direction as $\omega_D$ is
increased. In the regions of phase space which compose the three
phases identified only for the case $\omega_D=0$, the density profiles
approach continuously, as $\omega_D$ is decreased towards zero, to the
ones described in the corresponding phases of the $\omega_D=0$ case.

\section{Monte-Carlo simulations}
\label{sec:mc} 
\begin{figure}[t]
\begin{flushleft}
\includegraphics*[angle=270,width=5cm]{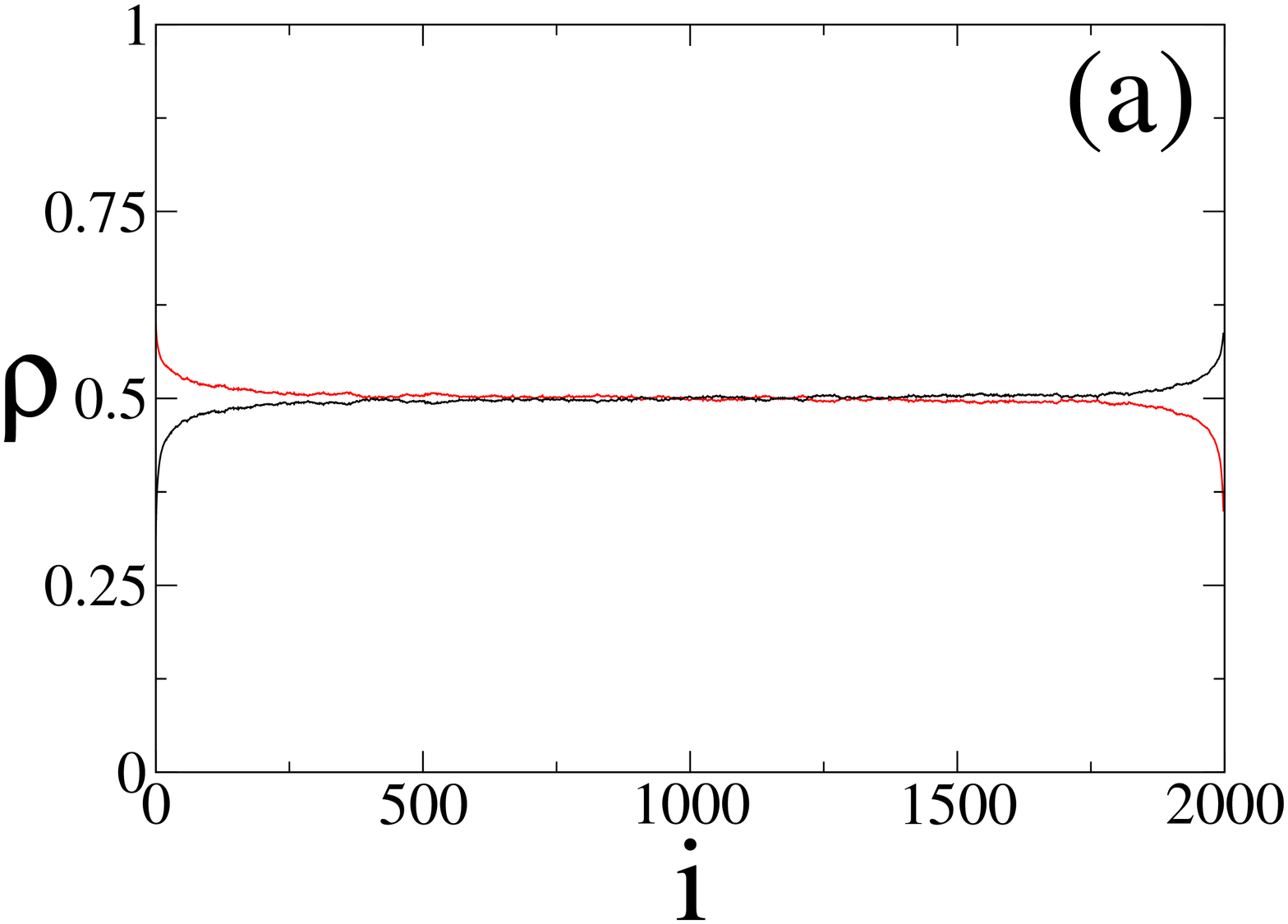}
\includegraphics*[angle=270,width=5cm]{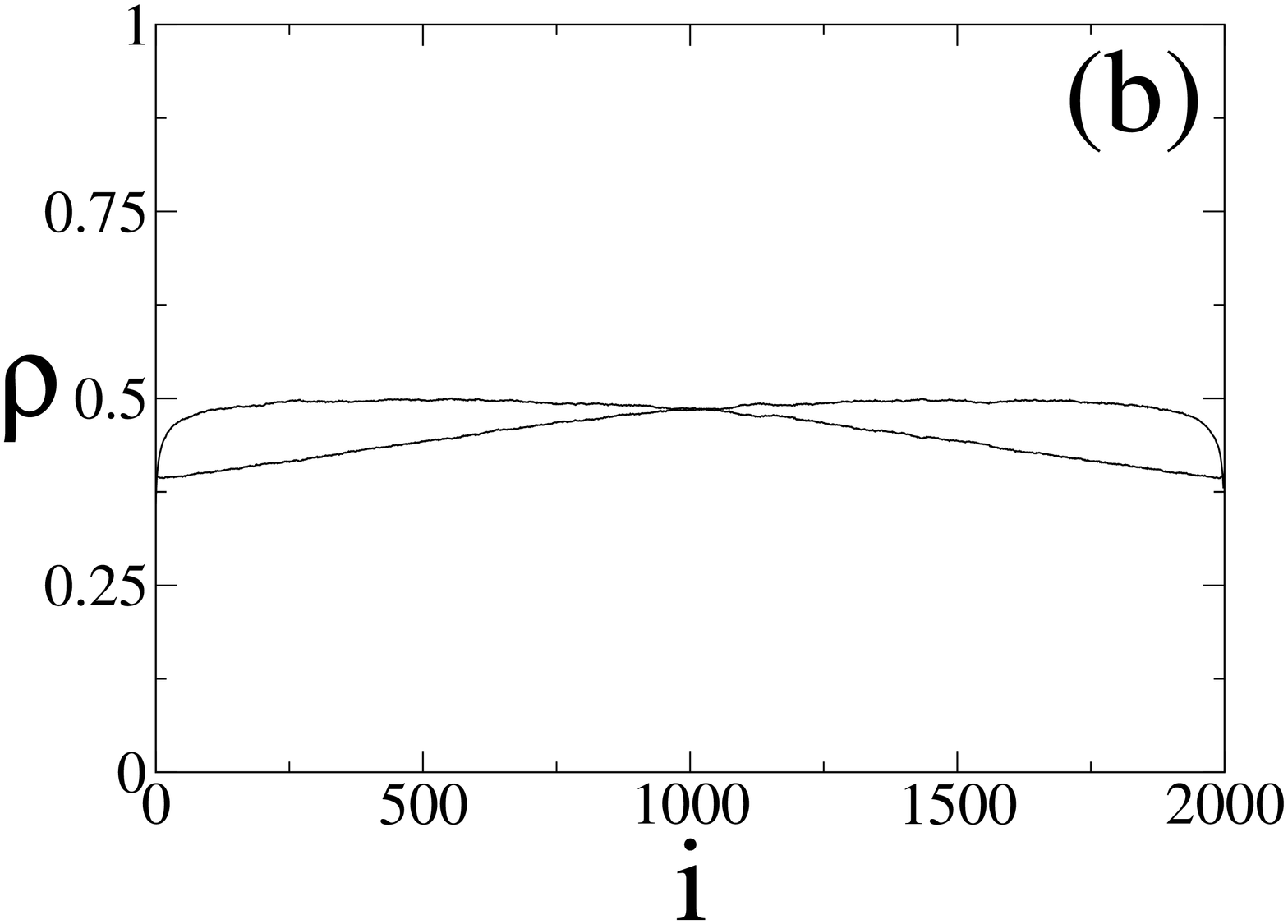}
\includegraphics*[angle=270,width=5cm]{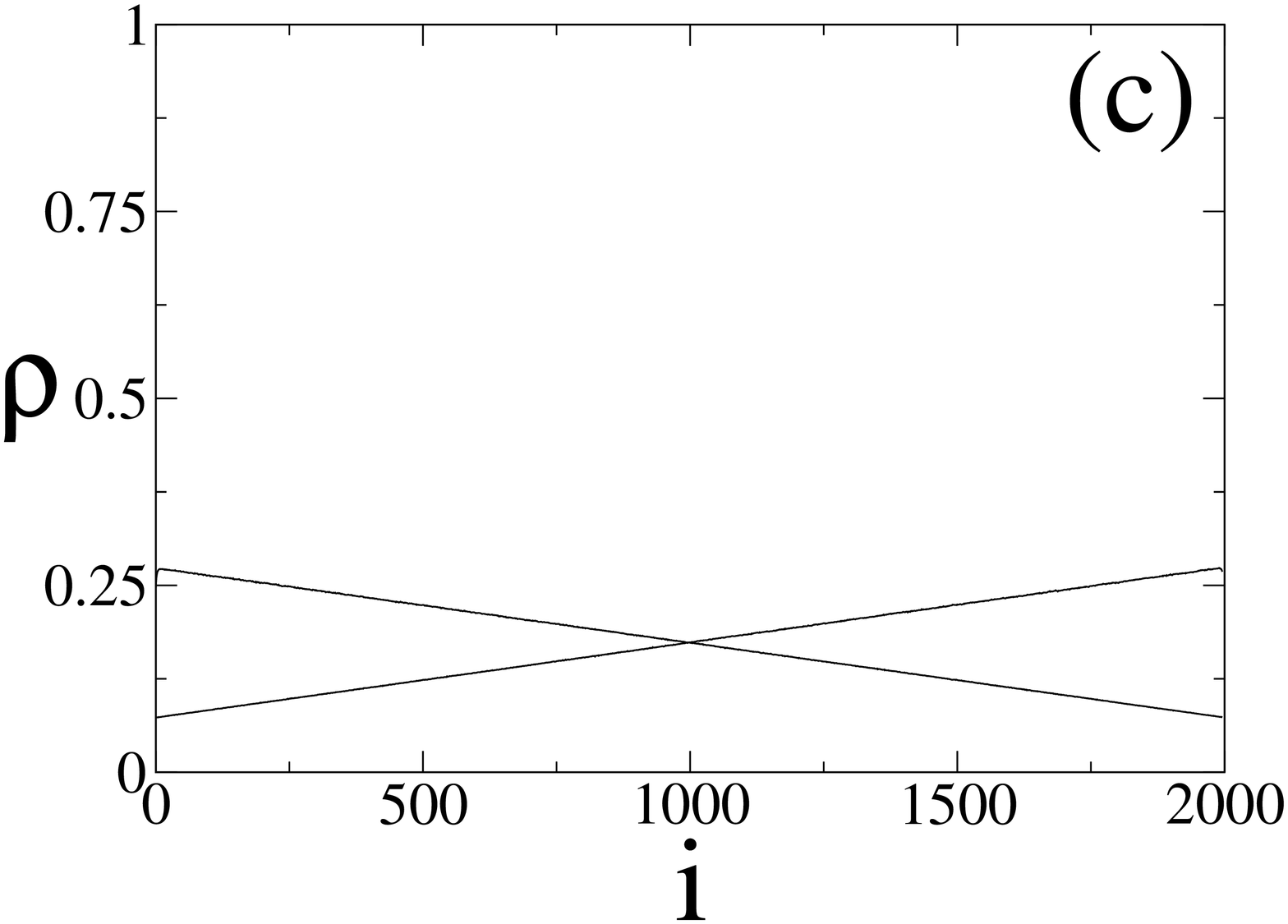}
\\
\includegraphics*[angle=270,width=5cm]{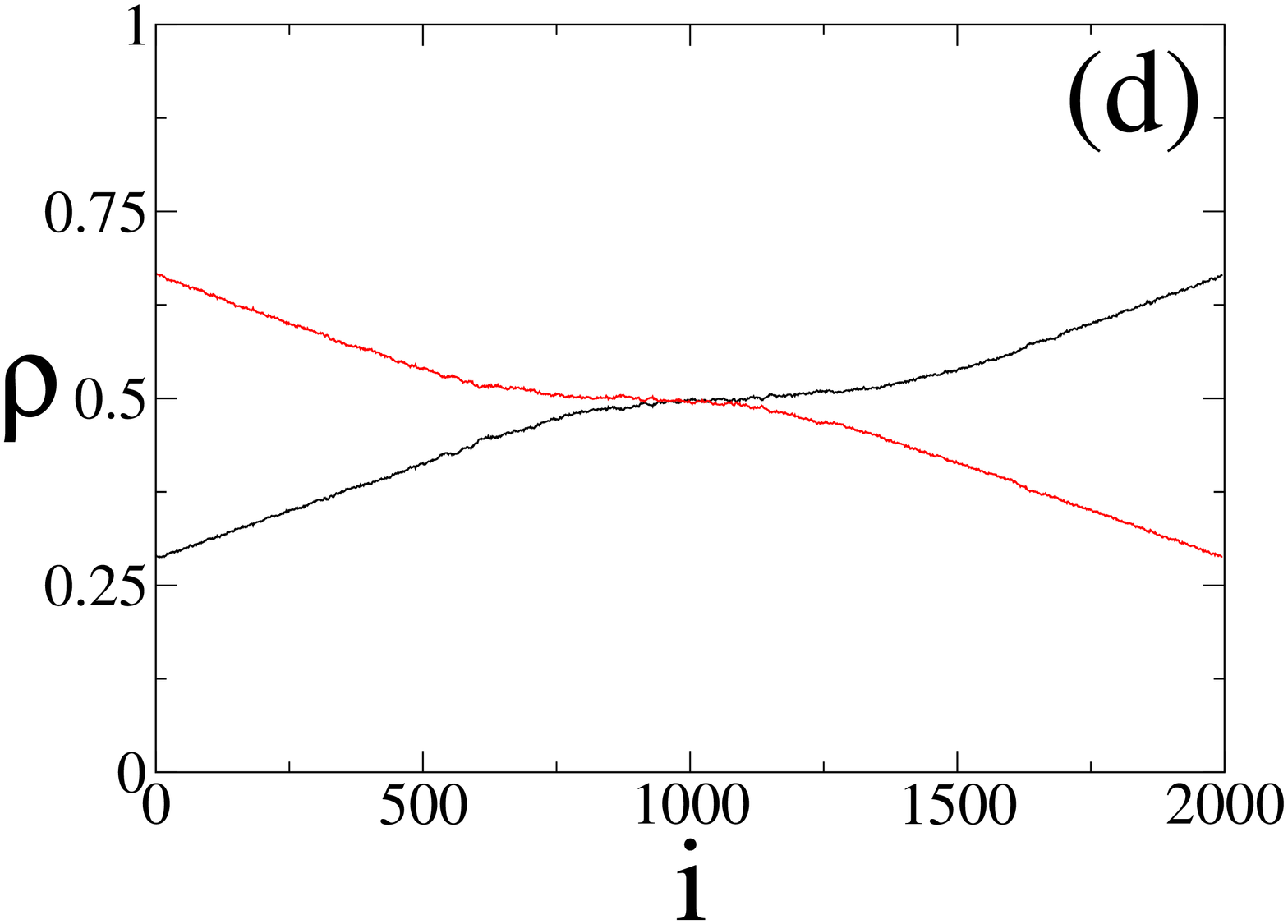}
\includegraphics*[angle=270,width=5cm]{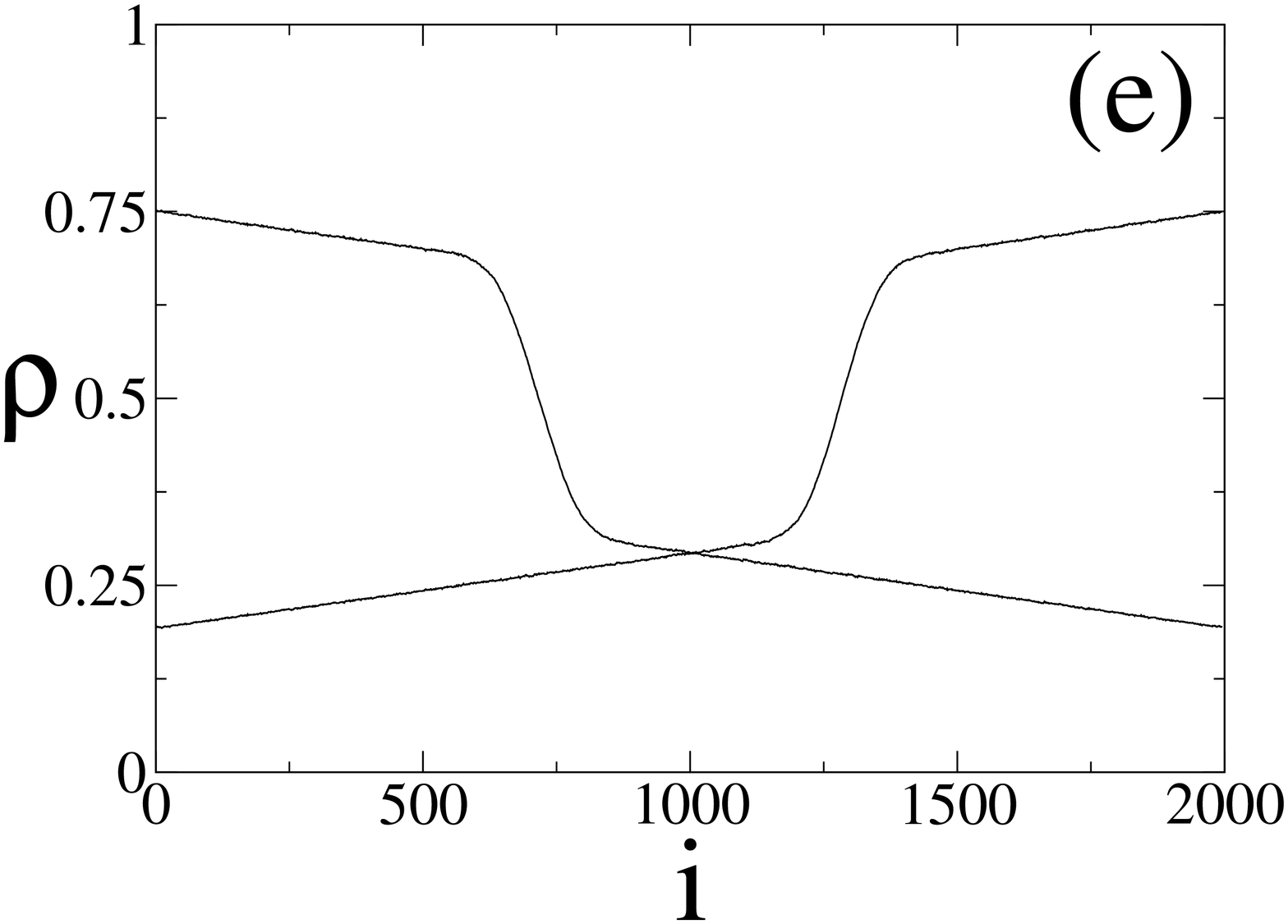}
\includegraphics*[angle=270,width=5cm]{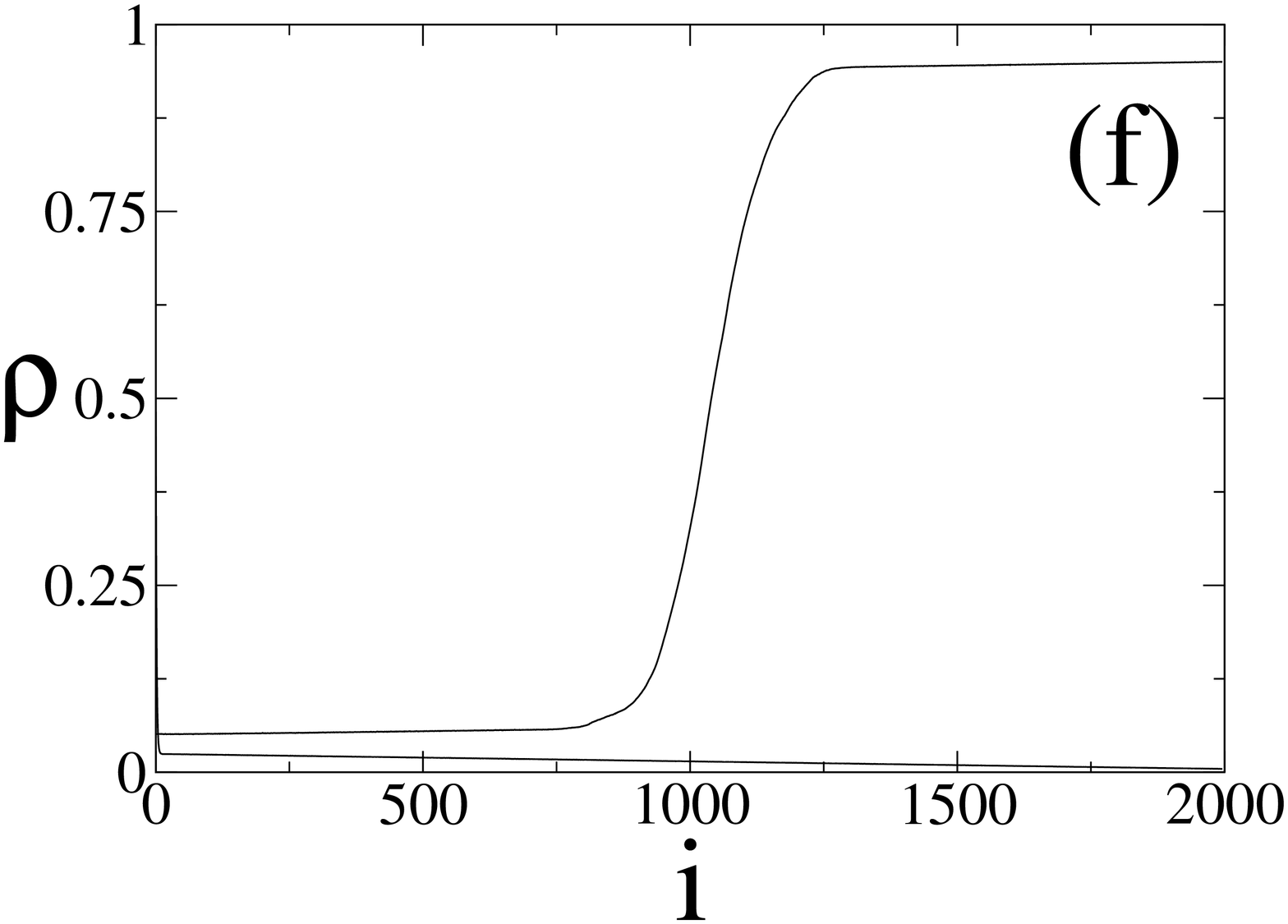}
\\
\includegraphics*[angle=270,width=5cm]{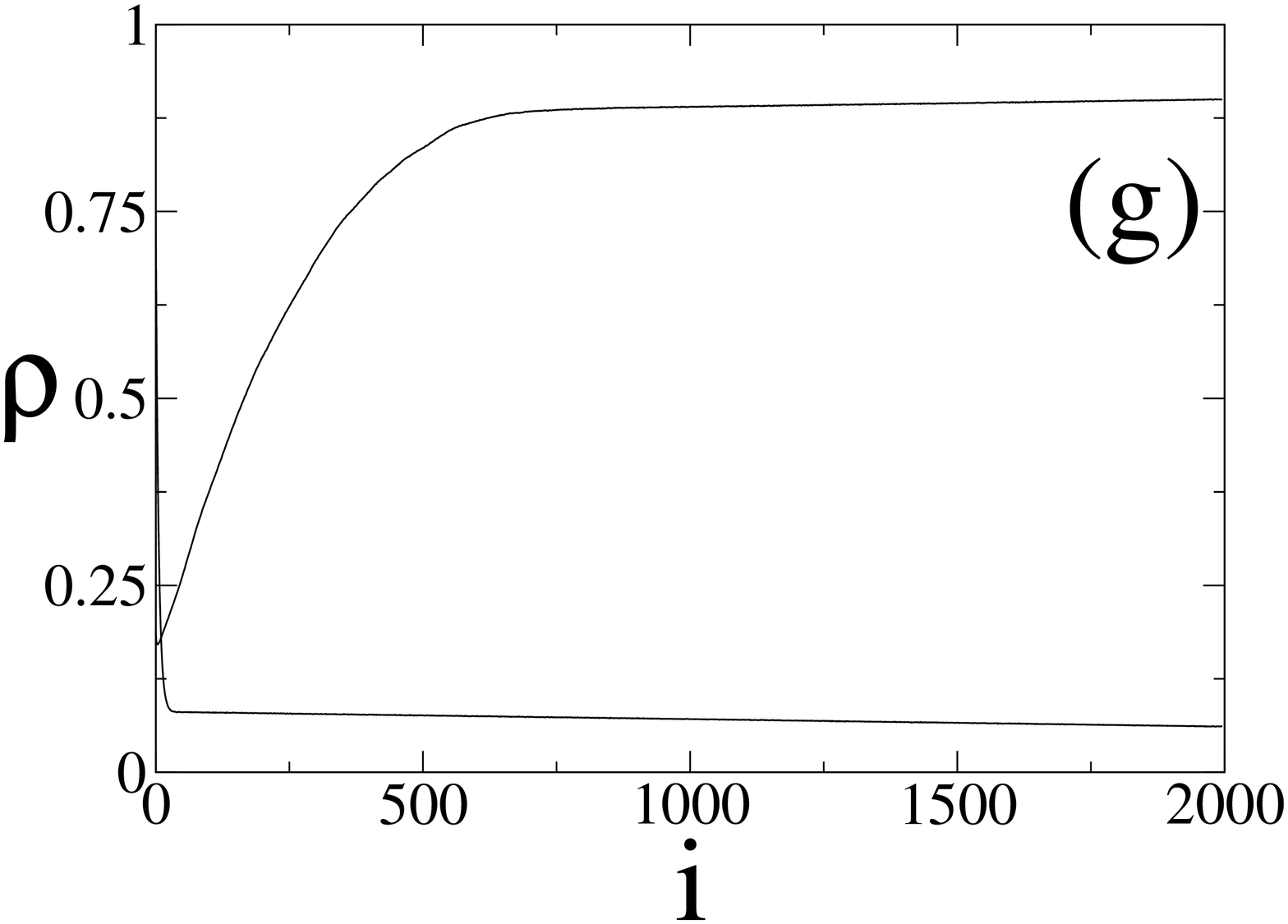}
\caption{Density profiles as obtained from a Monte-Carlo
simulation of a system of size $N=2000$. (a) Max Phase,
$\alpha=3.0$, $\beta=0.8$, $\Omega=0.2$, $q=1$. (b) Low-Max Phase,
$\alpha=1.0$, $\beta=0.7$, $\Omega=0.2$, $q=1$. (c) Low Phase,
$\alpha=0.1$, $\beta=0.8$, $\Omega=0.2$, $q=1$. (d) Low-Max-High
Phase, $\alpha=5.0$, $\beta=1/3$, $\Omega=0.5$, $q=1$. (e) Shock
Phase, $\alpha=3.0$, $\beta=0.25$, $\Omega=0.2$, $q=1$. (f)
Shock-Low asymmetric Phase, $\alpha=0.1$, $\beta=0.05$,
 $\Omega=0.02$, $q=1$.
(g) High-Low asymmetric Phase, $\alpha=1.5$, $\beta=0.1$,
$\Omega=0.02$, $q=1$. }
\label{fig:MCprofiles}
\end{flushleft}
\end{figure}

As was already noticed in earlier works about the
Bridge model \cite{Evans95,Arndt98}, the mean field phase diagram
captures the correct topology of the phase boundary lines. The
exact location of the boundary lines, however, is shifted in the
noisy model with regard to those of the mean field solution. For
our model we did not try to obtain the exact location of the phase
boundary lines from Monte-Carlo (MC) simulations. Still we note,
based on our simulations, that these lines cannot lie too far from
those of the mean-field phase diagram obtained in the previous
section. Here we concentrate on giving evidence for each of the
phases by finding representative points in parameter space. In
figure \ref{fig:MCprofiles} we present the density profiles of the
two species in the different phases as obtained from MC
simulations. The profiles in each phase were taken at the same
parameters as the respective mean field profiles shown in figures
\ref{fig:MFsymm} and \ref{fig:MFasymm}. For all phases shown here
the mean field profiles capture the features of the noisy model.
While MC simulations were done for a system of size $N=2000$, the
mean field results are obtained for $N=128$. A quantitative
comparison between the profiles becomes meaningful in the limit $N
\to \infty$ using rescaled coordinates $x=i/N$. In this limit the
localized shocks in the Shock-Low and symmetric Shock phase become
sharp \cite{Parmeggiani03}. In figure \ref{fig:MCprofiles} the
density profile for the Low asymmetric Phase is omitted. This
phase is addressed in the following subsection.


%
%
\subsection{The Low asymmetric phase}
\label{sec:lowlow} The existence of the Low asymmetric phase is an
issue of longstanding discussion \cite{Arndt98,Clincy01}. It was
noted already in the mean-field solution of the Bridge model
\cite{Evans95} that the region in phase space covered by this
phase is very small compared to the others. Furthermore MC
simulations indicate that the particle densities in this region of
phase space fluctuate strongly \cite{Clincy01}. These facts also
hold true in the model considered here. Therefore we refrain from
presenting a MC density profile as was done for the other phases.
In \cite{Clincy01} the existence of the Low-asymmetric phase in
the Bridge model is deduced from sampling the probability
distribution $P(\bar{p},\bar{m})$, where $\bar{p}$ and $\bar{m}$
are the average densities of positive and negative particles in
the system respectively. In our model, this line of argument
fails. On the level of average densities, the Low-Low and
Shock-Low phases cannot be distinguished. This is because both
phases can exhibit distributions $P(\bar{p},\bar{m})$ with two
peaks at $\bar{p}$ and $\bar{m}$ smaller than $\half$.

The blockage picture outlined in section \ref{sec:pic} yields no
arguments in favor of the Low asymmetric phase. At the upper
boundary of the Shock-Low phase the localized shock position
retracts to $x_s=0$. This allows particles of both species to
enter the system. A symmetry broken Low phase beyond this point
would require some kind of blockage being formed at the exit of
the majority species. The nature of such a blockage is not clear.
It remains to be clarified, whether the existence of this phase,
which is evident in the mean field treatment, can be demonstrated
in the noisy model.

\section{Induced shocks}
\label{sec:induced}

In the case $q \neq 1$ the bulk dynamics of the two particle
species are not decoupled. Thus, one cannot solve the mean field
equations in the way it was done in section \ref{sec:mft}. Still,
the phase structure can be explored by integrating the mean field
equations numerically, or by MC simulations. It shall not be
attempted here to give the full phase diagram of the model. We do
note, however, that phases with broken symmetry exist also in the
general case.

\begin{figure}
\includegraphics[angle=270,width=7.5cm]{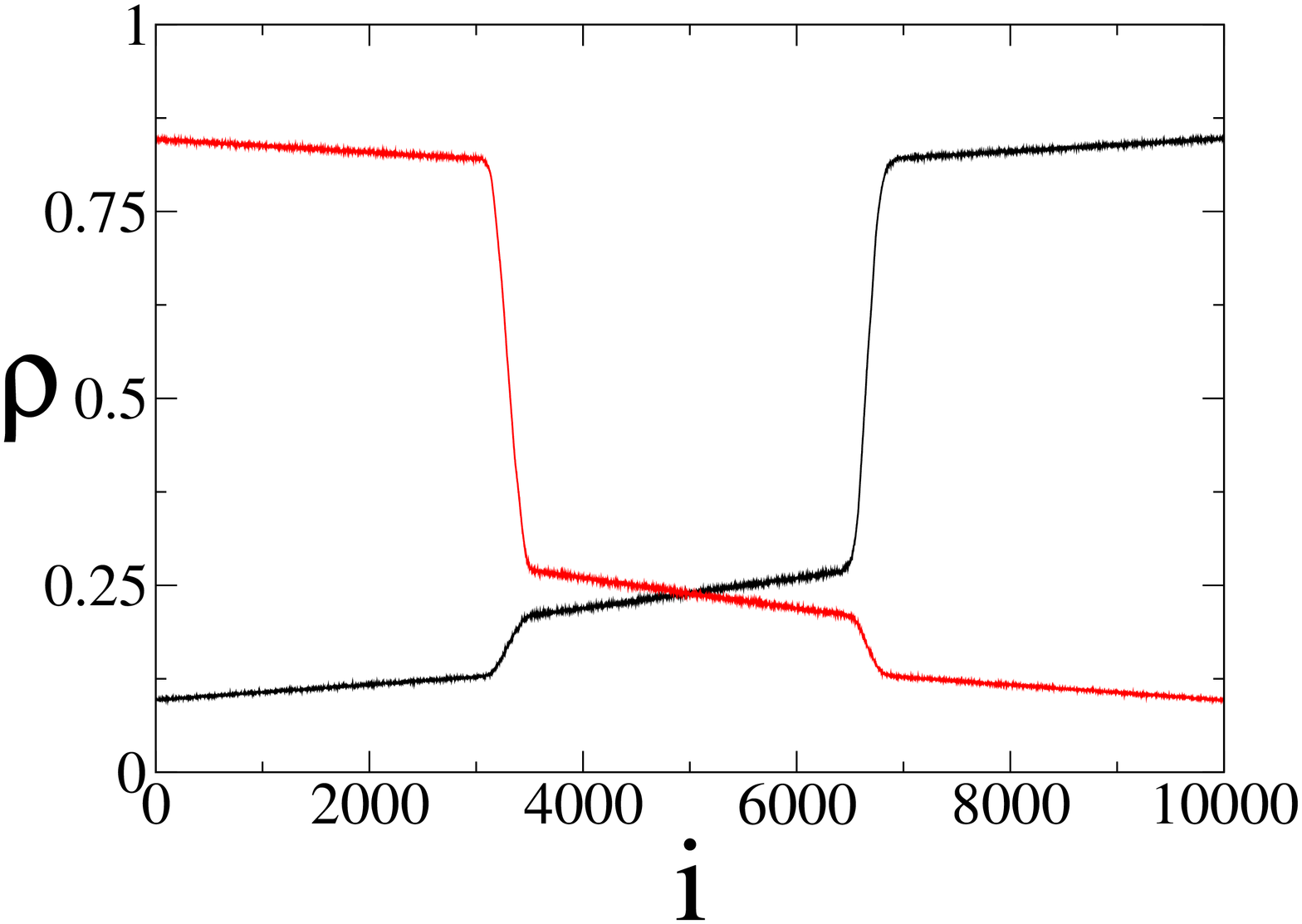}
\includegraphics[angle=270,width=7.5cm]{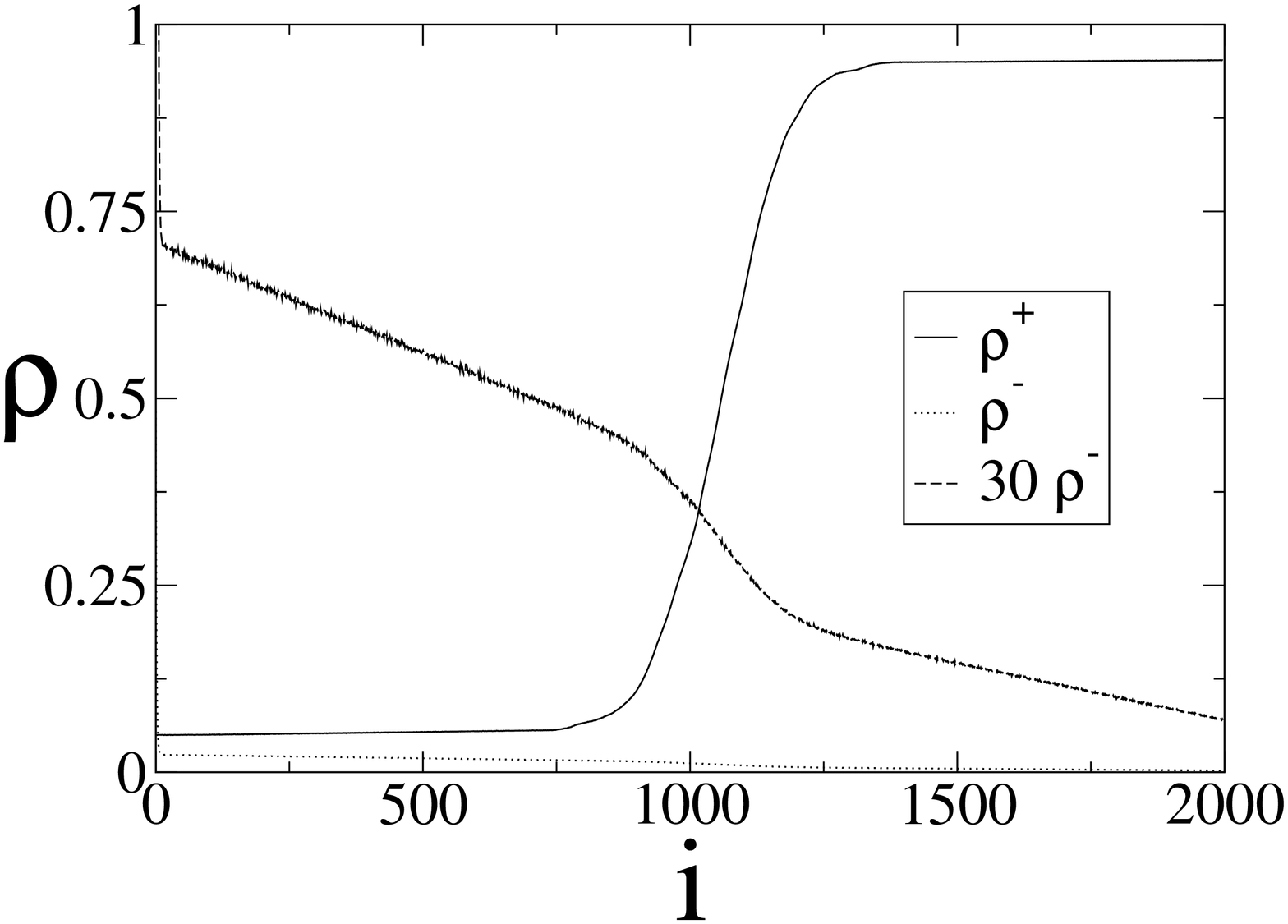}
\caption{\label{fig:induced} {\em Induced shocks.} (Left) Density
profiles in the symmetric shock phase, as obtained in Monte-Carlo
simulations. A primary shock in one of the species is accompanied
by an induced shock in the other species. Here $q = 2$ and
$\alpha = 3.0$, $\beta = 0.25$, $\Omega = 0.02$. (Right) Density
profiles in the Shock-Low phase, as obtained in Monte-Carlo
simulations. The localized shock in the majority species is
accompanied by an induced shock in the minority species. The
profile of the minority species is additionally shown when
multiplied by a factor of $30$, in order to demonstrate the
induced shock phenomenon. Here $q = 2$ and $\alpha = 0.$1, $\beta
= 0.05$, $\Omega = 0.02$. }
\end{figure}

In the Shock symmetric phase and the Shock-Low phase the coupling
of the dynamics of the two species gives rise to an {\em induced
shock} phenomenon. Here the existence of a localized shock in the
density profile of one species induces a shock in the density
profile of the other species. For example, in the Shock symmetric
phase one notices that the density profile of each one of the
species exhibits actually {\it two} shocks in the steady state
(figure \ref{fig:induced}). One is a {\em primary shock}, created
by the same localization mechanism which was already identified in
the bulk-decoupled case. The existence of this shock, albeit not
its detailed properties, relies only on the properties of the
density profile of the very same species. The second shock is {\em
induced} by the primary shock in the density of the other species,
and it shares its location. Across both shocks the current is
continuous. This phenomenon also occurs in the Shock-Low phase
(figure \ref{fig:induced}). Here, the localized shock in the
majority phase induces a shock in the Low phase. In fact, it was
shown in \cite{Popkov03A} that for general two species systems
with coupled density-current relation a shock in one particle
species induces a shock in the other one.

For general $q$ the current-density relation of, say, the positive particle species, $j_i^+(p_i,m_i)$,
depends on the local density of both species. In general this current-density relation is
not known, except for two cases: the case $q=1$ (the decoupled case), where $j^+=p(1-p)$, and
the case $q=2$, where $j^+=p(1-p+m)$ \cite{Evans95}.

As discussed in \cite{Popkov03B} the current across a localized
shock is continuous. This requirement implies when $q=1$ that
shocks are symmetric with respect to $p=1/2$, irrespective of the
local density $m$ of particles of the opposite species. In
general, however, this is not the case. The properties of both the
primary and induced shock in the density profile of each species
rely on the local densities of both.

The continuity of the current across the shocks can in principle be used to determine the properties of the primary and induced shocks,
if one can develop the density characteristics from the boundaries of the system.
For the case $q=1$ the equations for the two density profiles are decoupled, and one uses this method to determine the position of the shock.
Of course, no induced shocks are present in this case.
For general $q$, however, the equations for the density profiles are coupled, and an analytical solution is not available.

\section{Exact solution for the limit $\beta, \Omega \to 0$}
\label{sec:toy} In \cite{Godreche95} a {\em toy model} was
presented to allow for the exact solution of the limit
$\beta \to 0$ of the Bridge model. In this section a
generalization of the toy model is presented. The solution of this
model gives an exact description of the $\beta \to 0, \Omega \to
0$ limit of the model, and proves that to lowest order in $\beta$
mean-field theory recovers the exact phase diagram.

In the limit $\beta \to 0, \Omega \to 0$ the only relevant
configurations are those composed of three blocks, containing
(from left to right) negative particles, vacancies, and positive
particles. A configuration of this type is long-lived, as all exit
rates from it scale to zero. In this limit, all other
configurations rearrange themselves into one of these three-block
configurations. A configuration of this type is identified by two
variables, $y$ and $x$, defined as the size of the left (negative
particle) block and the right (positive particle) one,
respectively. Thus, for example,
\begin{equation*}\label{eq:toyblocks}\begin{array}{lcr}
\overbrace{----}^{N_-}\overbrace{0\;0\;0\;0\;0}^{N-N_--N_+}
\overbrace{+++++}^{N_+}&\longleftrightarrow&(x=N_+,y=N_-)\;.
\end{array}\end{equation*}

Let us assume that the system is in a three-block configuration
$(x,y)$, and consider the ways it can leave it. First, a
particle can leave the system through a boundary with rate $\beta$,
leaving a vacancy behind it. This vacancy can travel into the system
with rate $(1+\alpha)^{-1}$, in which case the system is
again in a three-block configuration. Otherwise, the particle which
had left the system can be replaced by a particle of the opposite
species with rate $\alpha(1+\alpha)^{-1}$. On a short time
scale this particle travels through the system until it joins the
block on the other side, thus returning the system into a three-block
state.

Another possible way out of a three-block configuration is through the
non-conserving processes in the bulk. First, a particle can be
attached to the system in the vacancy domain with rate
$\omega_A(N-x-y)$. This particle joins
on a short time scale to the block of its own species. Second, a positive
(negative) particle can be detached from the system with rate
$\omega_D x$ $(\omega_D y)$, thus creating a vacancy
within a particle block. On a short time scale this vacancy travels
into the system and joins the vacancy block. Finally, a positive
(negative) particle can change its species with rate $\omega_X x$
$(\omega_X y)$, and move from one particle block to the other.

When the last particle of its species leaves the system, the other
type of particles can rush into the system through the boundary. The
system fills rapidly with particles of this type. Thus, the only
possible configurations with $x=0$ or $y=0$ are $(0,N)$ and $(N,0)$.

\begin{figure}
\epsfysize 5 cm \epsfbox{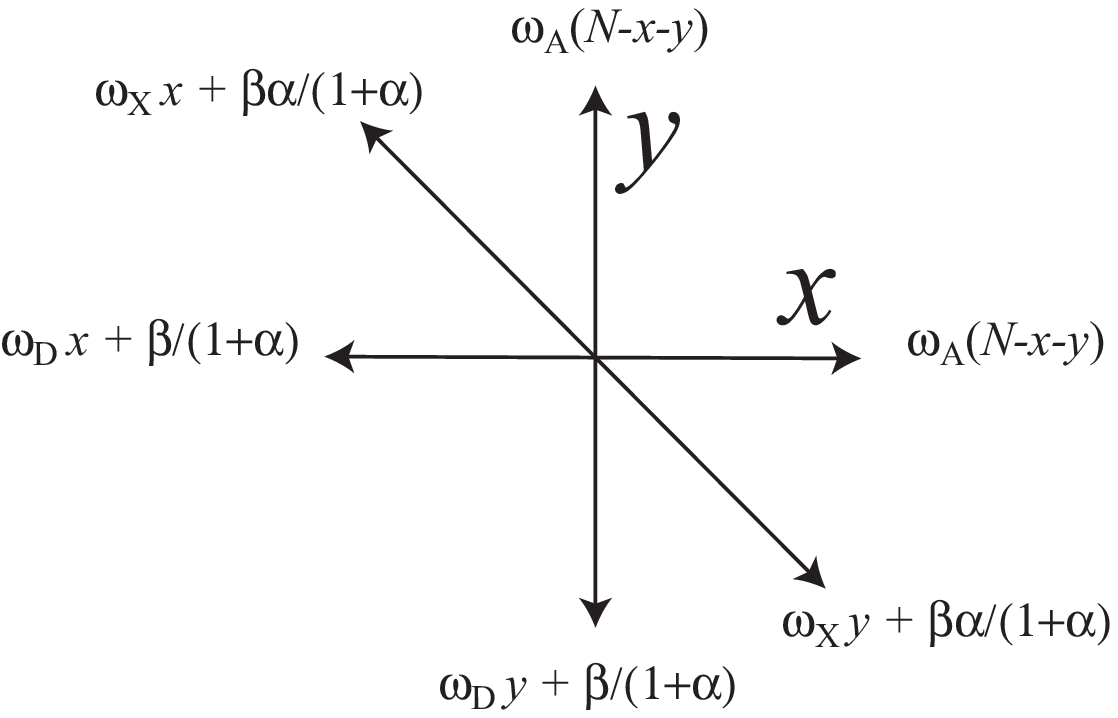} \epsfysize 5 cm
\epsfbox{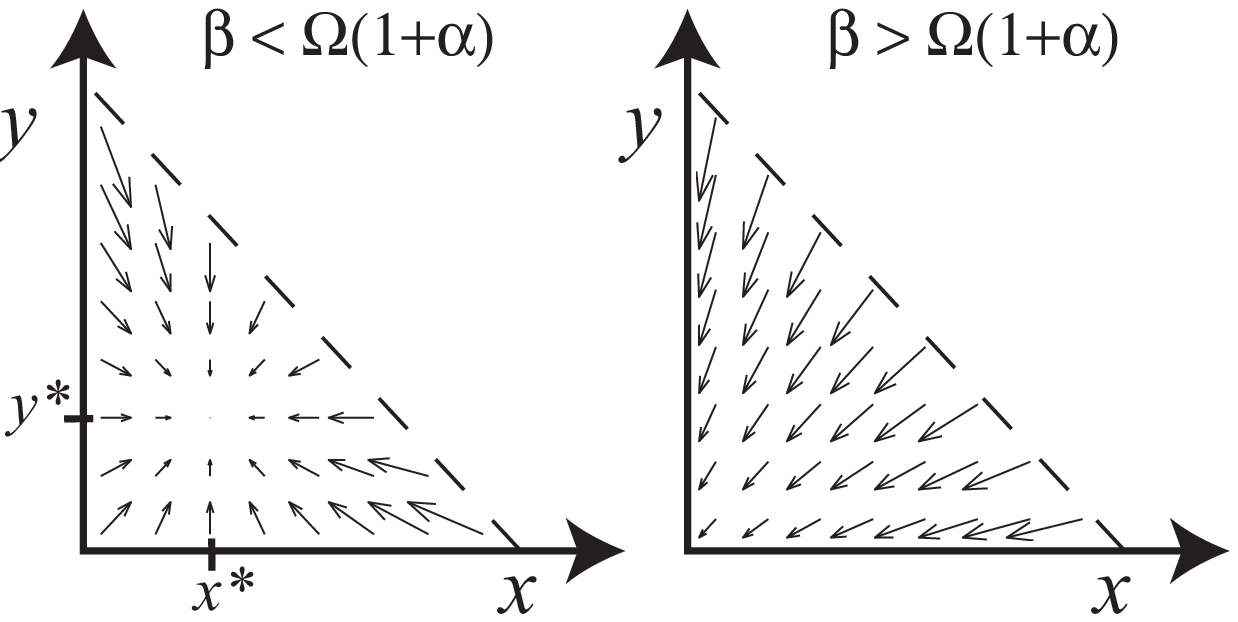} \caption{\label{fig:toy} {\em Toy model} for
the limit $\beta, \Omega \to 0$. (Left) The rates defining the
corresponding random-walk model (\ref{eq:toyrates}). (Right) Flow
fields of the model.}
\end{figure}

To summarize, consider a two-dimensional random walker, whose
position $(x,y)$ corresponds to the block-configuration of the
two-species system. The transition rates for this walker are (see
also figure \ref{fig:toy})
\begin{eqnarray}
\label{eq:toyrates}
(x,y) \to (x+1,y)\;\; &\mbox{ with rate }& \;\; \omega_A(N-x-y)
  \nonumber \\
(x,y) \to (x,y+1) \;\; &\mbox{ with rate }& \;\; \omega_A(N-x-y)  \nonumber \\
(x,y) \to (x-1,y) \;\; &\mbox{ with rate }& \;\; \beta\frac{1}{1+\alpha}
+ \omega_D{x}    \nonumber \\
(x,y) \to (x,y-1) \;\; &\mbox{ with rate }& \;\; \beta\frac{1}{1+\alpha}
+ \omega_D{y}    \\
(x,y) \to (x-1,y+1) \;\; &\mbox{ with rate }& \;\; \beta\frac{\alpha}{1+\alpha}
+ \omega_D{x}    \nonumber \\
(x,y) \to (x+1,y-1) \;\; &\mbox{ with rate }& \;\; \beta\frac{\alpha}{1+\alpha}
+ \omega_D{y}    \nonumber \\
(1,y) \to (0,N) \;\; &\mbox{ with rate }& \;\; \beta \nonumber \\
(x,1) \to (N,0) \;\; &\mbox{ with rate }& \;\; \beta \nonumber \; .
\end{eqnarray}
It can be shown that this
toy model is mapped exactly to the two-species model in the limit
$\beta, \Omega \to 0$, , in the sense formulated in appendix A of
\cite{Godreche95}.

Let us first consider the case where the dynamics defined in
(\ref{eq:toyrates}) leads to a fixed point solution. This is the case
where the net flows on both the $x$ and $y$ directions vanish at some
point $(x^*,y^*)$. The fixed points must satisfy the equations
\begin{eqnarray}
(\omega_X-\omega_A)x^* - (\omega_X+\omega_A+\omega_D)y^* &=&
-\omega_A
  N + \frac{\beta}{1-\alpha} \nonumber \\
- (\omega_X+\omega_A+\omega_D)x^* + (\omega_X-\omega_A)y^* &=&
-\omega_A
  N + \frac{\beta}{1-\alpha}\;,
\end{eqnarray}
whose solution is
\begin{equation}
\label{eq:toyfixedpt}
\frac{x^*}{N} = \frac{y^*}{N} = \frac{1}{2+u_D}\left(1-\frac{\beta}{\Omega(1+\alpha)}\right)\;.
\end{equation}
Interestingly, the fixed point does not depend on the charge exchange
rate $\omega_X$. Notice that this fixed point can only exist if
$0 \leq x^*/N, y^*/N \leq 1$. Indeed, $x^*$ and $y^*$ of
(\ref{eq:toyfixedpt}) always meet the second
condition. The first condition, however, is only met for
\begin{equation}
\label{eq:toyboundary}
\beta < \Omega(1+\alpha)\;.
\end{equation}
Otherwise, the random walker is always biased towards the axis of
the $(x,y)$ plane which is closer to its position (see figure
\ref{fig:toy}). In this case, $\Omega < \beta \to 0$, one recovers
the toy model of \cite{Godreche95} which yields a stable state of
broken symmetry. One of the species then occupies most of the
lattice, corresponding to the High-Low asymmetric phase of the
model.

Thus, the toy model yields two phases in the limit $\beta, \Omega
\to 0$. For $\beta < \Omega(1+\alpha)$ one has a symmetric phase,
with a dominating three-block configuration described by
(\ref{eq:toyfixedpt}). This corresponds to a symmetric shock phase
in the model, where the shock position of the positive particles
is $x_s = 1-x^*/N$, with $x^*$ given by (\ref{eq:toyfixedpt}).
Otherwise the system is in the High-Low asymmetric phase, with the
line $\beta = \Omega(1+\alpha)$ serving as the transition line
between the two phases.

For the bulk-decoupled case, $\omega_A = \omega_X$ and $q=1$, it is
illuminating to compare these exact results with the ones obtained
by mean field. The mean field analysis, performed in section
\ref{sec:mft}, predict in the limit $\beta, \Omega \to 0$  the two
phases obtained in the toy model. The mean field transition line
between the two phases (\ref{eq:conds}) is identical, to first
order in $\beta$, to the line $\beta = \Omega(1+\alpha)$ of the
toy model. Also the shock position $x_s$ calculated in the toy
model is identical to first order in $\beta$ to the one
(\ref{eq:shockposition}) calculated in mean field. This result
also holds in the case $\omega_D \neq 0$. We thus conclude that the
mean-field solution is exact to first order in $\beta$.

\section{Blockage picture}
\label{sec:pic}

In this section we combine the mean-field and stability analysis,
the simulation results, and the toy model into a physical picture.
Following \cite{Clincy01} we term it the {\em blockage picture}.

\begin{figure}
\begin{center}
\epsfbox{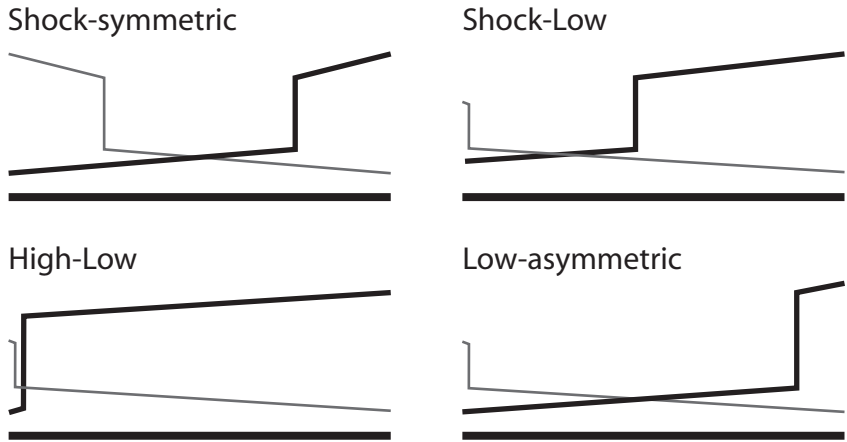}
\end{center}
\caption{\label{fig:blockage}  {\em Blockage picture.} Schematic
picture of the instantaneous density profiles. Density profiles of
positive particles are depicted by dark lines, negative particles
by gray lines. Here, as in the text, we assume that the positive
particles are the majority species.}
\end{figure}

Qualitatively, typical configurations in the asymmetric phases can
be described in terms of blocks of the two species, which spread
from the `exit' boundary into the system (figure
\ref{fig:blockage}, and compare the toy model description in the
previous section). The density profile within each block is not
constant, but this feature is not relevant here. As mentioned
above, a block of one species stalls the entry of particles of the
other species through the boundaries, thus serving as a blockage.
The possibility of particles to enter the system in the bulk
serves to stabilize the domain size.

In the High-Low phase the block of the majority species covers the
entire system, while in the Shock-Low phase the block is limited
to some part of the system. The fluctuations in the size of this
block, corresponding to the width of the localized shock, are
limited to an area of size $\sim N^{-1/2}$ \cite{Parmeggiani03}.
The minority block in both phases is unstable in the sense that
the domain wall between it and the bulk region drifts towards the
boundary. Averaging over the positions of the domain wall results
in the exponential decay of the mean field density profile from
the left boundary.

In \cite{Clincy01} it was observed that an instantaneous
configuration in the low-density asymmetric phases comprises a
small block of the majority phase, which is limited to the
vicinity of the boundary. The formation of this block prevents
particles of the other species from entering the system, thus
leading to symmetry breaking. However, this block does not survive
for times which are exponential in the system size.

The block picture is extended into the symmetric shock phase.
Here the two blocks are covering equal distances from the `exit' boundaries.
The sizes of the two blocks are again macroscopic and localized, in the sense that the size fluctuations
vanish as $N^{-1/2}$.

We now turn to describe the different phases of the model in the language of block configurations.
To this end we take a stroll along a line of constant $\alpha$ in the phase diagram, starting from the
symmetric shock phase and going up in $\beta$. This line is chosen such that it cuts through all
asymmetric phases.

In the shock symmetric phase, the two blocks inhibit, in a
symmetric way, the inflow of particles of the opposite species.
Increasing $\beta$ decreases the sizes of the two blocks. However,
as long as their size is macroscopic, the blocks keep their role
of lessening the ability of particles of the other species to
enter the system.

As the boundary line is approached, the size of the blocks is
reduced to zero. Now the road is open for both species to enter
the system. Due to fluctuations, the formation of a block is
inevitable. As $\beta$ is increased beyond the transition line
into the Shock-Low phase, the possibility rises that this block
will be stabilized by the non-conserving dynamics. A spontaneously
created block of one species, which now has a stable macroscopic
size, hinders particles of the other species and breaks the
symmetry between the two species.

As $\beta$ is increased from its value at the boundary line
between the symmetric shock phase and the Shock-Low phase, the
size of the block of the majority species increases. This is due
to the coupling between the ejection rate $\beta$ and the
effective injection rates, which at this region of phase space
serves to increase $\alpha^{+}$ (in mean-field this can be seen
from Eqs. \ref{eq:effective}, \ref{eq:bcsl}). At some value of
$\beta$, the block reaches the size of the entire system, and
there it stays for some range of $\beta$. This, in fact, is the
High-Low phase of the system. As $\beta$ is increased further the
size of the majority block shrinks back, and the system is again
in the Shock-Low phase. The transition from the Shock-Low phase to
the High-Low phase at some $\beta$, and the re-entrance to the
Shock-Low phase at some higher $\beta$, occur at these points
where the size of the majority blockage becomes identical with the
size of the system.

Towards the boundary line between the Shock-Low phase and the
low-density phase the size of the majority block vanishes. The
existence of a reminiscent block which yields the asymmetric Low
phase, as discussed in \cite{Clincy01}, can be either attributed
to fluctuations of the localized shock, or to an alternative
mechanism.

\begin{figure}
\begin{center}
\epsfxsize 15 cm \epsfbox{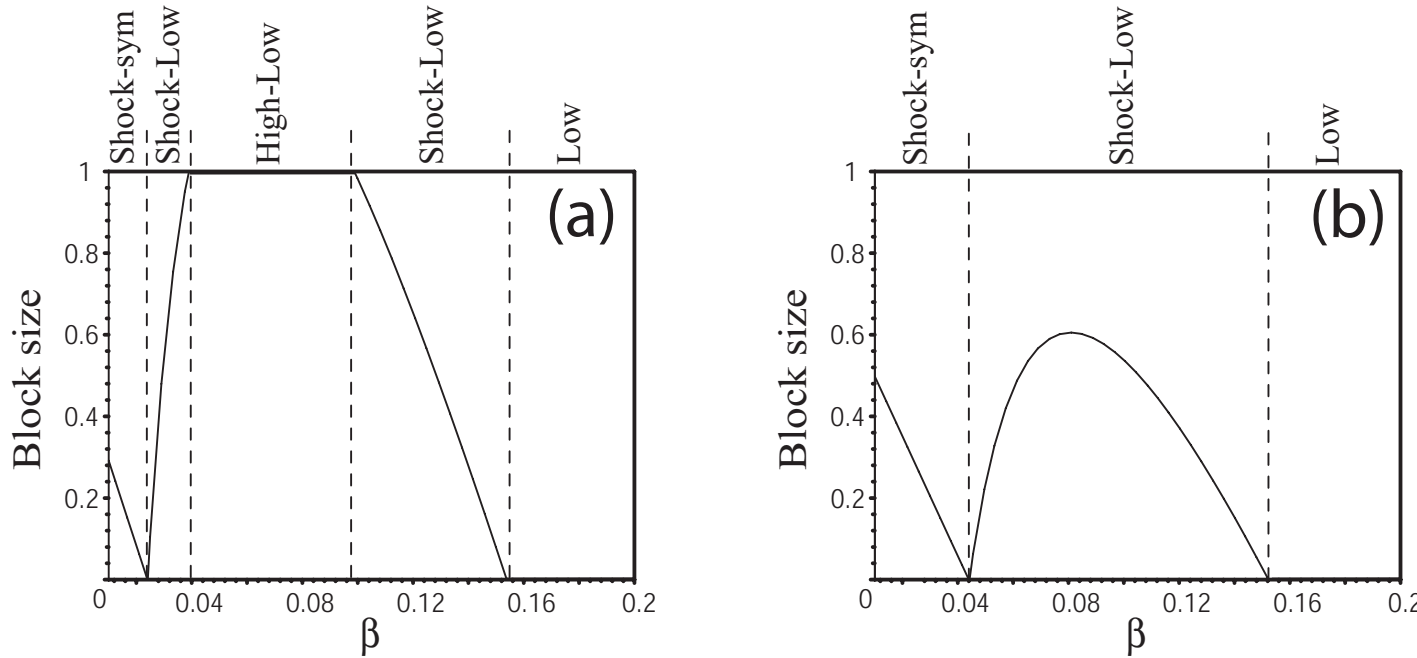}
\end{center}
\caption{\label{fig:xs} Blockage size as a function of $\beta$, as
calculated in mean-field for the bulk-decoupled case. Here
$\alpha=0.2, \omega_D=0$ and (a) $\Omega=0.02$, (b) $\Omega=0.03$.
}
\end{figure}

To make the blockage picture more quantitative, we calculate the
size of the majority block in each phase. This is done within
mean-field for the bulk-decoupled case. For simplicity we take
$\omega_D = 0$, where the profiles of the blocks are linear. The
block size is then just $1-x_s$, where the shock position $x_s$ is
given in (\ref{eq:shockposition}). In figure \ref{fig:xs} (a) we
plot the size of the block as a function of $\beta$ at constant
$\Omega$ and $\alpha$. Using the picture described above, one can
identify the phase boundaries. The value of $\beta$ at which the
blockage first disappears corresponds to the Shock-symmetric to
Shock-Low line. The two values between which the blockage spans
the system are identified as the two lines between the Shock-Low
and High-Low phases. Finally, the higher $\beta$ at which the
blockage disappears completely corresponds to a transition into a
low-density phase. It is easy to verify, by comparing with the
mean-field phase diagram, that these are indeed that transition
points between the phases. Note that the Low asymmetric phase
escapes this picture.

It is also possible to describe in terms of the block size (or
alternatively the shock position) the fact that asymmetric phases
disappear as $\Omega$ is increased. The non-conserving dynamics in
the bulk of the system serves to sustain the localized shock.
Keeping $\alpha$ constant, for example, the increase in the
amplitude of the non-conserving rates drives the position of the
shock out of the system, thus decreasing the maximal size of the
blockage. In terms of phases, this would decrease the segment on
the $\beta$ axis in which the shock is localized at the `entry'
boundary (\ie the High-Low phase), down to a point where the shock
cannot get so far and the phase disappears. Beyond this point, as
depicted in figure \ref{fig:xs} (b), the position of the shock is
driven back towards the `exit' boundary, thus reducing the
Shock-Low phase until it is finally gone.

\section{Summary}
\label{sec:summary}

In this work a two-species one-dimensional model, with dynamics
which is not conserving both at the boundaries and in the bulk,
has been studied.
The dynamics is symmetric under charge exchange and left-right reflection.
By definition the non-conserving dynamics in
the bulk of the system acts to diminish the difference between the
densities of the two species. Nevertheless we have found that the
symmetry between the two species can be broken even in the
presence of bulk non-conserving processes.

The mean-field phase diagram, obtained for the case where the bulk
dynamics of the two species becomes decoupled, exhibits three
phases in which the symmetry between the species is broken. One of
these is unique to the case where the bulk dynamics is not
conserving, and results from a localization of shocks in the
density profiles. All asymmetric phases reside in regions of phase
space where symmetric low-density profiles are another fixed point
of the mean-field dynamics. However, stability arguments shows
that it is the asymmetric solution which should be expected to
survive fluctuations. Indeed, comparing with Monte-Carlo
simulations, two asymmetric phases are confirmed. The third, in
which the average density of both species is below $\half$ is more
difficult to determine.

In contrast to the bulk-conserving case, in this model the density
profiles are generally not flat. In particular, localized shocks
may be generated in the bulk of the system. In the general case,
when the particle current of one species depends on the density of
the other, a localized shock in the density profile of one species
induces a shock in the other.

In the asymmetric phases, as well as in the shock symmetric phase,
typical configurations can be described in terms three blocks. The
leftmost block has a high density of negative particles, the
middle block has a low density of particles of both species, and
the right block is mainly occupied by positive particles. This
observation serves to define a toy model which describes the
dynamics of the system in terms of a two-dimensional random
walker. Solving the toy-model yields an exact solution for the
case where the exit rates and the non-conserving rates are taken
to zero. The results coincide with the ones obtained in mean-field
at this limit. For the general case a more qualitative picture
emerges, which serves to describe the phase transitions in the
model in terms of the block sizes.

The bulk of this work, as well as of those which studied the
Bridge model, has focused on the case in which the dynamics of the
two species is decoupled in the bulk. The other, more general
case, was studied only by numerical means, both on the mean-field
level and in Monte-Carlo simulations. This enabled us to observe
induced shocks. A more detailed study of this case by analytical
means should shed more light on this phenomenon.

The mean-field phase diagram of the NC-TASEP is expected to be
exact \cite{Popkov03B}, while Monte-Carlo results suggest that
this is not the case here. Such is also the case in the Bridge
model. It should be interesting to study the correlations
which build up in the system, taking it away from the mean-field
description.

\ack
It is a pleasure to thank D. Mukamel, and G.M. Sch\"utz  for
suggestions, discussions and critical reading of the manuscript.
We thank M.R. Evans, and M. Paessens, and V. Rittenberg for useful
discussions. The Einstein Center and the Forschungszentrum
J\"ulich are gratefully acknowledged for support during mutual
visits.
The support of the Israeli Science Foundation is gratefully acknowledged.

\appendix
\section{Mean-Field analysis of the case $\omega_D > 0$}
In this Appendix we discuss the construction of the phase diagram
for the bulk-decoupled case $q=1, \omega_A = \omega_X$ with
$\omega_D \neq 0$. This case corresponds to the NC-TASEP with
non-equal attachment and detachment rates (the case $K \neq 1$ in
\cite {Parmeggiani03,Popkov03B,Evans03}). The phase diagram of the model in the case
$\omega_D \neq 0$ exhibits two symmetric phases, symmetric
low-density phase and shock symmetric one, and three asymmetric
ones, High-Low density phase, Shock-Low phase, and low-density
asymmetric phase. The profiles do not have constant slopes as in
the $\omega_D=0$ case. Starting from left boundary density
$\alpha$, the density at the right boundary resulting from the
left characteristic is given by
\begin{equation}
\label{eq:rhol} \rho_{N}^{(\ell)}(\alpha)=\frac{2-u W_0
\left(\frac{(2-4\alpha-2\alpha u)}{u} \exp\left({-{\frac
{4\,\alpha+2\,\alpha\,u+\Omega(2+u)^2-2}{u}}}\right) \right)}
{2(2+u)}.
\end{equation}

The respective expression for the density at the left boundary
resulting from the right characteristics starting from density
$1-\beta$ reads
\begin{equation}
\label{eq:rhor} \rho_{1}^{(r)}(\beta)= \frac{2-uW_{-1}\left(
\frac{(4\beta-2-2u+2\beta u)}{u}
\exp{\left(\frac{\Omega(2+u)^2-2(1+u)+2\beta (2+u))}{u}\right)}
\right)}{2(2+u)}\;.
\end{equation}

Our aim is to use the known phase diagram of the NC-TASEP
\cite{Parmeggiani03} to construct the phase diagram of the two-species
model, as it was done in section \ref{sec:wd0}. To this end, let
us define the two transition lines in this phase diagram. The
first is the transition line between the high density phase and
the localized shock phase,
\begin{equation}
\label{eq:hsline}
 \hspace{-1.0cm}\beta_{\rm HS}(\alpha)=\frac{
2(1+u)+u {\rm W}_0\!\left(\frac{(4\alpha-2-2u+2\alpha u)}{u}{
\exp\!\left({\frac{-4\alpha+2+2u-2\alpha u+\Omega (2+u)^2 }{u}}\right)}\right)}{2(2+u)}\;.
\end{equation}
where ${\rm W}_k(z)$ is the Lambert-W function. This equation
defines the line only for $\alpha < 1/2$. The second transition
line separates the shock phase and the low density phase. This
line is given, for  $\alpha < 1/2$, by
\begin{equation}
\label{eq:slline} \beta_{\rm SL}(\alpha)=\frac{ 2-u {\rm
W}_{-1}\!\left(\frac{(2-4\alpha-2\alpha u)}{u}{
\exp\!\left({-{\frac
{4\,\alpha+2\,\alpha\,u+\Omega(2+u)^2-2}{u}}}\right) } \right) }
{2(2+u)}\;.
\end{equation}
Both boundary lines are continued for $\alpha>1/2$ by $\beta_{\rm
HS}(\half)$ and $\beta_{\rm SL}(\half)$, respectively.

 Let us recall the procedure in which one obtains the
phase diagram for the two-species model. For each phase, one
obtains from (\ref{eq:effective}) the effective boundary rates,
$\alpha^+$ and $\alpha^-$. This requires knowledge of the four
boundary currents, $j^+_b = p_b(1-p_b)$ and $j^-_b = m_b(1-m_b)$,
where $b=0,N$ for the left and right boundaries. The boundary
lines are then obtained from comparing the effective boundary
rates with the corresponding transition lines of the NC-TASEP.

We do not repeat the analysis here in such details as it was done
for the case $\omega_D=0$. Instead, we give for each phase the
four boundary densities, needed to calculate the boundary currents
and the effective rates. In addition the conditions on the
effective rates, which define the phase boundaries, are given in
terms of $\beta_{\rm HS}$ and $\beta_{\rm SL}$ of Eqs.
(\ref{eq:hsline}) and (\ref{eq:slline}). The asymmetric
low-density phase is omitted, as we could not obtain the boundary
densities in this phase in a closed form. The boundary lines for
this case were obtained numerically.

{\it Low density symmetric phase.} Here, $p_0=m_N=\alpha^+$,
$p_N=m_1=\rho_{N}^{(\ell)}(\alpha^+)$ as in (\ref{eq:rhol}). The
condition for the existence of this phase is

\begin{equation}
\beta<\beta_{\rm SL}(\alpha^+)\;.
\end{equation}

{\it Shock symmetric phase.} The boundary densities are given by
$p_0=m_N=\alpha^+$, $p_N=m_1=1-\beta$. The conditions for this
phases existence are
\begin{equation}
\beta_{\rm SL}(\alpha^+) < \beta  <  \beta_{\rm HL}(\alpha^+)\;.
\end{equation}

{\it Shock - Low asymmetric phase.} Let the positive particles be
in the shock phase. Then $p_0=\alpha^+$, $p_N=1-\beta$. The
negative particles are in the Low phase, where $m_N=\alpha^-$ and
$m_1=\rho_{N}^{(\ell)}(\alpha^-)$ as given by equation
\ref{eq:rhol}. The phase exists in a region in phase space where
\begin{equation}
\beta_{\rm SL}(\alpha^+) < \beta  <  \beta_{\rm SL}(\alpha^-)\;.
\end{equation}

{\it High - Low asymmetric phase.} As before it is assumed that
the positive particles are in the majority phase:
$p_0=\rho_{1}^{(r)}(\beta)$ according to \ref{eq:rhor} and
$p_N=1-\beta$. The negative particles are in the Low phase:
$m_N=\alpha^-$ and $m_1=\rho_{N}^{(\ell)}(\alpha^-)$. The High-Low
phase exists where
\begin{equation}
\beta_{\rm HL}(\alpha^+) < \beta  <  \beta_{\rm SL}(\alpha^-) \;.
\end{equation}

\end{document}